\documentclass[11pt]{article}
\usepackage{amsmath}
\usepackage{latexsym}
\usepackage{color}
\usepackage{graphics, graphicx}
\usepackage{amsmath,amssymb,mathrsfs}
\newtheorem{thm}{Theorem}

\newtheorem{cor}[thm]{Corollary}

  \def\ulamek#1#2{\mbox{\normalfont$\frac{#1}{#2}$}}

\numberwithin{equation}{section}

\def\be{\begin{eqnarray}}
\def\ee{\end{eqnarray}}

\let\text=\textstyle

\def\bE{\mathbb E}

\def\be{\begin{eqnarray}}
\def\ee{\end{eqnarray}}
\def\ben{\begin{eqnarray*}}
\def\een{\end{eqnarray*}}
\def\bei{\begin{itemize}}
\def\eei{\end{itemize}}

\begin{document}

%[Explicit representations for multiscale L\'{e}vy processes]
\title{Explicit representations for multiscale L\'{e}vy processes, and asymptotics of multifractal conservation laws.}

\author{K.~G\'{o}rska\footnote{H. Niewodnicza\'{n}ski Institute of Nuclear Physics, Polish Academy of Sciences, Division of Theoretical Physics, ul. Eliasza-Radzikowskiego 152, PL 31-342 Krak\'{o}w, Poland},
 and W.~A.~Woyczy\'{n}ski\footnote{Department of Mathematics, Applied Mathematics and Statistics, and Center for Stochastic and Chaotic Processes in Science and Technology, Case Western Reserve University, Cleveland, OH 44122, U.S.A.}}
%\email{katarzyna.gorska@ifj.edu.pl}

%\affiliation{H. Niewodnicza\'{n}ski Institute of Nuclear Physics, Polish Academy of Sciences, Division of Theoretical Physics, ul. Eliasza-Radzikowskiego 152, PL 31-342 Krak\'{o}w, Poland}

%\affiliation{Department of Mathematics, Applied Mathematics and Statistics, and Center for Stochastic and Chaotic Processes in Science and Technology, Case Western Reserve University, Cleveland, OH 44122, U.S.A.}

%\email{waw@case.edu}

%\affiliation{Department of Mathematics, Applied Mathematics and Statistics, and Center for Stochastic and Chaotic Processes in Science and Technology, Case Western Reserve University, Cleveland, OH 44122, U.S.A.}

%\pacs{..}

\maketitle

\begin{abstract}

Nonlinear conservation laws driven by L\'evy processes have solutions  which, in the case of supercritical nonlinearities,   have an asymptotic behavior dictated by the solutions of the linearized equations. Thus the explicit  representation of the latter is of interest in the nonlinear theory. In this paper we concentrate on the case where the driving L\'evy  process is a multiscale  stable (anomalous) diffusion,   which corresponds to the case of multifractal conservation laws considered in \cite{BFW1999, BKW1999,BKW2001,BKW2001a}. The explicit representations, building on the previous work on single-scale problems (see, e.g.,\cite{KGorska11}),  are developed in terms of the special functions (such as  Meijer G functions), and are amenable to    direct numerical evaluations of relevant probabilities.

\end{abstract}

%---------------------------------  Section 1 -------------------------------------------

\section{Introduction}

Mathematical conservation laws are integro-differential evolution equations, such as Navier-Stokes and Burgers equations,  expressing the physical principles of conservation of mass, energy, momentum, enstrophy, etc., in different dynamical situations. With this paper we initiate a program of investigation of explicit representations for asymptotic behavior of solutions of conservation laws driven by multiscale, $(\alpha_1,\dots,\alpha_k)$-stable,  L\'{e}vy  processes (multifractal anomalous diffusions). Since the asymptotic behavior of such conservation laws is determined, in some cases, by their linearized versions, the starting point here is to obtain  exact representation, via known special functions such as Meijer G functions, of the solutions of linear multiscale evolution equations, that is, for the PDFs of the multiscale L\'evy processes themselves.

The idea is to produce a framework that permits a straightforward calculation of probabilities related to the mutiscale diffusions using a symbolic manipulation platform such as {\it Mathematica},  and a fairly standard set of special functions that have been in use in this area for a long time.

The plan of the paper is  as follows. We begin, in Section 2, with a  review of the known general results from  \cite{BFW1999, BKW1999, BKW2001, BKW2001a} on the asymptotics of solutions of multifractal conservation laws, and apply them to the case of general asymmetric two sided multiscale diffusion. In the case of   supercritical nonlinearity  asymptotics is dictated by the linear part of the equation so, in Section 3,  we  produce an exact representation of   solutions of linearized equations in the general asymmetric case; simpler representation are then deduced in the symmetric case. In Section 4, anticipating our future needs to obtain explicit solutions for equations describing subdiffusive anomalous diffusions, where the time is also "fractal" \cite{PSW2005}, we obtain an explicit representation for totally asymmetric $\alpha$-stable diffusions, for $0 < \alpha < 1$. Conclusions, as well as a discussion   of the relevant moment problem,  can be found in Section 5.

%\subsection{Preliminaries and notations}

In the remainder of this section we establish the  notation and provide basic definitions of the integral transforms and special functions we are going to work with. We start with the Fourier transform of an integrable function $f(x)$ defined for $x \in \mathbb{R}$, and real $\omega$ \cite{INSneddon72},
\begin{align}\label{26/08/14-1a}
\tilde{f} (\omega) &= \mathcal{F}[f (x); \omega] = \int_{-\infty}^{\infty} e^{i \omega x} f (x) dx,\\ \label{26/08/14-1b}
f (x) &= \mathcal{F}^{-1}[\tilde{f} (\omega); x] = \frac{1}{2\pi} \int_{-\infty}^{\infty} e^{-i \omega x} \tilde{f} (\omega) d\omega.
\end{align}
For a function $f (x) \equiv 0$,  for $x < 0$, such that  $ e^{- c x} f (x)$  is integrable on the positive half-line for some fixed number $c > 0$, the Laplace transform  
\begin{align}\label{26/08/14-2a}
f^{\star} (p) &= \mathcal{L}[f (x); p] = \int_{0}^{\infty} e^{-p x} f (x) dx, \\ \label{26/08/14-2b}
f (x) &= \mathcal{L}^{-1}[f^{\star} (p); x] = \frac{1}{2\pi i} \int_{c-i \infty}^{c+i \infty}e^{p x} f^{\star} (p) dp,
\end{align}
where $p = c + i \omega$. There is an obvious relationship between the Fourier transform of $f_{1}(x)  = e^{- c x} f_{2}(x)$  and the Laplace transform of $f_2(x)$, see e.g.,  \cite{INSneddon72} for more information. Finally,  the Mellin transform of $f (x)$ is here defined as follows:
\begin{align}
\label{26/08/14-3a}
\hat{f} (s) &= \mathcal{M}[f (x); s] = \int_{0}^{\infty} x^{s-1} f (x) dx,\\ 
\label{26/08/14-3b}
f (x) &= \mathcal{M}^{-1}[\hat{f} (s); x] = \frac{1}{2\pi i} \int_{L} x^{-s} \hat{f} (s) ds, 
\end{align}
where $s$ is a complex variable \cite{INSneddon72}. The contour of integration $L$ is determined by the domain of analyticity of $\hat{f}_{2}(s)$ and, usually,  it is an infinite strip parallel to the imaginary axis. The conditions under which the integrals in Eqs. \eqref{26/08/14-1a}-\eqref{26/08/14-3b} converge can be found in \cite{INSneddon72}.

The Meijer $G$ function  will play a pivotal role in what follows. It is defined as the inverse Mellin transform of products and ratios of the classical Euler's gamma functions. More precisely, see  \cite{APPrudnikov-v3, NIST},
\begin{equation}\label{26/08/14-4}
G^{m, n}_{p, q}\left(z\Big\vert {A_{1} \ldots A_{p} \atop B_{1} \ldots B_{q}}\right) = \mathcal{M}^{-1}\left[\frac{\prod_{j=1}^{m} \Gamma(B_{j} + s)\, \prod_{j=1}^{n} \Gamma(1 - A_{j} - s)}{\prod_{j=m+1}^{q} \Gamma(1 - B_{j} - s) \prod_{j=n+1}^{p} \Gamma(A_{j} + s)}; z\right],
\end{equation}
where  empty products in   Eq. \eqref{26/08/14-4} are taken to be equal to 1. Eq. \eqref{26/08/14-4} holds under the following assumptions:
\begin{align}
\label{26/08/14-5}
&z\neq 0, \quad 0 \leq m \leq q, \quad 0 \leq n \leq p, \nonumber\\
&A_{j}\in\mathbb{C}, \quad j = 1, \ldots, p; \quad B_{j}\in\mathbb{C}, \quad j = 1, \ldots, q.
\end{align}
A description of the integration contours in Eq. \eqref{26/08/14-4}, and the general properties and special cases of the Meijer $G$ functions can be found  in  \cite{APPrudnikov-v3}.  If the integral in Eq. \eqref{26/08/14-4} converges and if no confluent poles appear among $\Gamma(1 - A_{j} - s)$ or $\Gamma(1- B_{j} - s)$, then the Meijer $G$ function can be expressed as a finite sum of the generalized hypergeometric function, see formulas (8.2.2.3) and (8.2.2.4) on p. 520 of \cite{APPrudnikov-v3}.  Recall that a generalized hypergeometric function can be represented in terms of the following series, see Eq. (7.2.3.1) on p. 368 of \cite{APPrudnikov-v3}:
\begin{equation}\label{26/08/14-6}
{_{p}F_{q}}\left({ a_{1}, \ldots, a_{p} \atop b_{1}, \ldots, b_{q}}; x\right) = \sum_{n=0}^{\infty} \frac{x^{n}}{n!} \frac{\prod_{j=1}^{p} (a_{j})_{n}}{\prod_{j=1}^{q} (b_{j})_{n}},
\end{equation}
where  the upper and lower lists of parameters are denoted by $(a_{p})$ and $(b_{q})$, respectively, and $(a)_{n} = \Gamma(a + n)/\Gamma(a)$ is the Pochhammer symbol.

To conclude the introduction we find it convenient to introduce the special notation  for a specific uniform partition of the unit interval, 
\begin{equation}\label{speciallist}
\Delta(n, a)=\{ \ulamek{a}{n}, \ulamek{a+1}{n}, \ldots, \ulamek{a+n-1}{n}\}.
\end{equation}
For later reference, we also quote  the Euler's reflection formula, 
\begin{equation}\label{26/08/14-7}
\Gamma(z) \Gamma(1-z) = \frac{\pi}{\sin(\pi z)}, 
\end{equation}
see Eq. (8.334.3) on p. 896 in \cite{Gradshteyn}, 
and the Gauss-Legendre multiplication formula, 
\begin{equation}\label{26/08/14-8}
\Gamma(n a) = (2\pi)^{\frac{1-n}{2}} n^{n a - \frac{1}{2}} \prod_{j=0}^{n-1} \Gamma\left(a + \frac{j}{n}\right),
\end{equation}
see Eq. (8.335) on p. 896 of \cite{Gradshteyn}.

%-------------------------------------------   Section 2   ---------------------------------------------------------------
 
\section {Multiscale L\'evy processes} 

In this Section we are  turning  to a review of   infinitesimal generators  $\cal A$  of  semigroups associated with 1-D  multiscale $(\alpha_1,\dots,\alpha_k)$-stable L\'evy processes driving the  evolution equations of the form, 
\begin{equation}\label{29/12/2014-1}
\frac{\partial u}{\partial t}+{\cal A} u + \frac{\partial}{\partial x}g(u) =0, \qquad u(0, x) = u_{0}(x),
\end{equation}
where  $u=u(t,x), t\ge 0, x \in \mathbb R$,   $u_{0}$ is an initial condition which will be specified later, and $g:\mathbb R\mapsto \mathbb R$ is a (nonlinear) function. Such equations are often called fractal, or anomalous  conservation laws \cite{BKW1999,BKW2001}.  Their asymptotic behavior will be discussed in Section 3.

The  operators $\cal A$  are easiest to describe in terms of their actions in the Fourier domain; they are  so-called Fourier multiplier operators. Let us begin by recalling the basic terminology and establishing the notation. 
%\medskip
 
Like any Markov processes\footnote{See, e.g., \cite{JS2001}, for basic information in this area.}, the L\'evy process, $X_t, t>0$,  has associated with it a semigroup $P_t$ of convolution operators\footnote{ That is, $P_{t+s} = P_t P_s$, $t, s > 0$.} acting on a bounded function $\phi(x)$ via the formula,
 \begin{equation}\label{29/12/2014-2}
P_t \phi(x)=\bE ^x(\phi(X(t))=\int_{\mathbb R} \phi(x+y)\,P(X(t)\in dy).
 \end{equation}
The {\it infinitesimal generator} $\cal A$ of such a semigroup is defined by the formula,
 \begin{equation}\label{29/12/2014-3}
-{\cal A}=\lim_{h\to 0} {P_h-P_0\over h} ,
 \end{equation}
and the family of functions,$v(t, x)~=~P_t g(x)$,   interpreted here  as  probability density functions (PDFs),   clearly satisfies   the (generalized) Fokker-Planck evolution equation, 
\begin{equation}\label{29/12/2014-4}
\frac{\partial v}{\partial t} = -{\cal A} v, 
\end{equation}
because $ \lim\limits_{h\to 0} (P_{t+ h}-{P_t})/h = \lim\limits_{h\to 0}[({P_{ h}-P_0)/ h}]P_t =- {\cal A}P_t$.
   
In the case of a general L\'evy processes $X_t$, we have the identity,
 \begin{equation}\label{29/12/2014-5}
{\cal F}({\cal A} \phi)(\omega)= \psi (\omega){\cal F}\phi(\omega),
 \end{equation}
where $\cal F$ stands for the Fourier transform, and 
\begin{equation}\label{29/12/2014-6}
\psi (\omega)=\log \mathbb{E}[e^{i\omega X_1}]
\end{equation}
is the characteristic exponent of $X_1$, which is necessarily (see, e.g., \cite{B1996}), of the form 
\begin{equation}\label{29/12/2014-7}
\psi(\omega )= i\mu \omega - \frac{(\sigma \omega )^2 }{2} + \int_{\mathbb{R}}(e^{i \omega  x}-1-i\omega x  {\bf I}_{|x|<1})\Lambda(dx),
\end{equation} 
where $\mu \in \mathbb{R}$, $ \sigma \in \mathbb{R_{+}}$, and $\Lambda$  is a nonegative measure on $\mathbb R$, satisfying the conditions $\Lambda(\{0\})=0$,  and $\int_{\mathbb{R}}(1 \wedge |x|^2)\Lambda(dx) <  \infty$. The triplet $(\mu,\sigma,\Lambda)$ is called the \textit{characteristic triplet} of $X$, $\mu\in\mathbb{R}$ --  the drift coefficient, $ \sigma> 0 $ --  the Gaussian, or diffusion coefficient, and $ \Lambda $ --  the L\'evy measure of $X_1$. The L\'evy measure describes the ``intensity'' of jumps of a certain height of a L\'evy process  in a time interval of length 1.
Observe that 
\begin{align*}
{\cal F}(P_t \phi)(\omega) & = \left(\int_R e^{-i\omega x}\bE \phi(X_t+x)\,dx\right) =\bE \left(\int_Re^{-i\omega (y-X_t)} \phi(y)\,dy\right) \\
& = \bE e^{i\omega X_t}\int_Re^{-i\omega y} \phi(y)\,dy=\exp( t\psi( \omega )){\cal F} \phi(\omega ), 
\end{align*}
which,   in view of Eq. \eqref{29/12/2014-3},  indeed implies Eq. \eqref{29/12/2014-5}.  
 
In the case of the usual Brownian motion the infinitesimal operator ${\cal A} $ is just the 1-D classical Laplacian $\Delta$ (the second derivative operator). For the self-similar (single-scale) symmetric $\alpha$-stable process $X_t$, the infinitesimal generator is the 1-D fractional Laplacian {$-(-\Delta)^{\alpha/2}, \; 0 < \alpha \leq 2$,} corresponding to the characteristic exponent (Fourier multiplier) $\psi(\omega)=-|\omega|^\alpha$.

In what follows we focus our attention on the multiscale (and not necessarily symmetric) L\'evy processes with the characteristic functions of the form,  
\begin{equation}\label{30Jun14-2}
\bE e^{i\omega X_t}=\prod_{j=1}^n \tilde{v}_{\alpha_{j}, \beta_{j},\gamma_j}(\omega, t),  
\end{equation}
where, for each $j=1,2,\dots, n$, 
\begin{equation}\label{29/12/2014-8}
\tilde{v}_{\alpha_{j}, \beta_{j}, \gamma_{j}}(t, \omega)=   \mathcal{F}[v_{\alpha_{j}, \beta_{j}, \gamma_{j}}(t, x);\omega] = \exp\left[- t \gamma_j|\omega|^{\alpha_{j}} e^{\frac{i\pi}{2} \beta_{j}\, {\rm sgn}(\omega)}\right]
\end{equation}
The symbol ${\rm sgn}(\omega)$ denotes the sign of the parameter $\omega$. The  multiparameter  $(\vec \alpha;\vec \beta;\vec \gamma) = (\alpha_1,\dots,\alpha_n;\beta_1,\dots \beta_n;\gamma_1,\dots,\gamma_n)$ has to satisfy the following conditions:  If  $0 < \alpha_{j} < 1$ then   $|\beta_{j}| \leq \alpha_{j}$, and if $1 < \alpha_{j} \leq 2$ then  $|\beta_{j}| \leq 2 - \alpha_{j}$; for all $j$,  we assume that $\gamma_j > 0$ . Thus the Fourier multiplier describing the infinitesimal generator of $X_t$ is of the form,  
\begin{equation}\label{multisymbol}
\psi_{{(\vec \alpha;\vec \beta;\vec \gamma)} }(\omega)=\sum_{j=1}^n -  \gamma_j|\omega|^{\alpha_{j}} e^{\frac{i\pi}{2} \beta_{j}\, {\rm sgn}(\omega)}.
\end{equation}
The generator itself will be denoted ${\cal A}_ {(\vec \alpha;\vec \beta;\vec \gamma)}$. For the sake of convenience, and without loss of generality, in the remainder of the paper we will assume that  
$$
\alpha_1<\alpha_2<\dots<\alpha_n.
$$

The densities $v_{\alpha_{j}, \beta_{j}}(x, t)$ appearing in \eqref{29/12/2014-8} are unimodal \cite{ELukacs70, WFeller}. The skewness parameter,  $\beta_{j}$, measures the degree of  asymmetry of $v_{\alpha_{j}, \beta_{j}}(x, t)$: for $\beta_{j} = 0$  they are just  the  previously mentioned symmetric $\alpha_j$-stable densities  with fractional Laplacians as the corresponding infinitesimal generators.  Moreover, all of those densities are self-similar, since, for any $x\in\mathbb{R}$, and $  t > 0$,
\begin{equation}\label{31/08/14-1}
v_{\alpha_{j}, \beta_{j},\gamma_j}(t, x) = \frac{1}{t^{1/\alpha_{j}}} v_{\alpha_{j}, \beta_{j},\gamma_j}\left(1, \frac{x}{t^{1/\alpha_{j}}}\right),
 \end{equation}
and 
\begin{equation}\label{29/12/2014-8a}
v_{\alpha_{j}, \beta_{j}, \gamma_{j}}(-x, t) = v_{\alpha_{j}, -\beta_{j}, \gamma_{j}}(t, x).
\end{equation}
Eq. \eqref{29/12/2014-8a} is a consequence of the identity $\tilde{v}_{\alpha_{j}, \beta_{j}, \gamma_{j}}(t, -\omega) = \tilde{v}_{\alpha_{j}, -\beta_{j}, \gamma_{j}}(t, \omega)$ satisfied by $\tilde{v}_{\alpha_{j}, \beta_{j}, \gamma_{j}}(t, \omega)$ given in Eq. \eqref{29/12/2014-8}.

 %--------------------------------------------------    Section 3   ----------------------------------------------------------
 
\section {Asymptotics of solutions of  multifractal conservation laws   with  supercritical nonlinearity} 

Now, we are ready to state the results about   existence, uniqueness, and the asymptotic behavior of the solutions of the Cauchy problem for the multifractal conservation laws \eqref{29/12/2014-1}  driven  by multiscale L\'evy  processes (anomalous diffusions) introduced in Section 2. The main point here is the observation that the solutions of Eq. \eqref{29/12/2014-1},  under certain conditions on the generator $\cal A$ and the nonlinearity $g$,   have the large time behavior   similar to solutions of Eq. \eqref{29/12/2014-4}\footnote{This is in contrast to the phenomena observed for data of Riemann type (nonintegrable, and nonsmooth), when  shocks are created, see .e.g., \cite{DI2006, D2010,GW2014}.}. These results provide the  motivation for the   work presented in the following sections. {The physical justification for considering  conservation laws driven by 
L\'evy processes are more numerous than can be cited here, but see, e.g.,  \cite{WAW2001} for a review of the subject.}

The solutions of Eq. \eqref{29/12/2014-1}  have to be understood in some weak sense which opens several possibilities presented for example in \cite {BFW1999, BKW1999, BKW2001, BKW2001a, KW2008}. Motivated by the classical Duhamel formula we choose to interpret them as the so-called {\it mild} solutions satisfying the identity, 
\begin{equation}\label{D}
u(t, x) = \left[e^{ -t {\cal A}}u_0\right]\!(x) - \int_{0}^{t} \frac{\partial}{\partial x} e^{- (t-\tau){\cal A}}[g(u)](\tau, x) \,d\tau.
\end{equation}
 
The basic results are summarized in the following Theorem, where 
the regularity of the solutions of Eq. \eqref{D} is expressed in terms of the Sobolev space $W^{2,2}$.

%----------------------  Theorem 1 -----------------------
\begin{thm}\label{t3.1}
(see {\rm\cite{BKW2001}}) (i) Assume that $g\in C^1({\mathbb R}, {\mathbb R}^{d})$ and ${\cal A}$ is the infinitesimal generator of a L\'evy process with the symbol satisfying  the condition
\begin{equation}\label{infH}
\limsup_{|\omega|\to\infty}{{\psi(\omega)-\psi_0|\omega|^2} \over{|\omega|^{\widetilde\alpha}}}<\infty\ \ \mbox{for some } 0< \widetilde\alpha<2, {\;and}\; \psi_0>0.
\end{equation} 
Given $u_0\in L^1({\mathbb R} )\cap L^\infty({\mathbb R} )$, there exists a unique solution $u\in {\cal C}([0,\infty); \;L^1({\mathbb R} )\cap L^\infty({\mathbb R}))$ of the problem
\begin{equation}\label{L1}
\frac{\partial u}{\partial t} + {\cal A}u + \nabla\cdot g(u) = 0, \qquad u(x,0)=u_0(x).
\end{equation}
This solution is regular, $u\in C((0,\infty)$; $W^{2,2}({\mathbb R}))\cap C^1((0,\infty); L^2({\mathbb R} ))$, satisfies the conservation of integral  property, $\int u(x,t)\, dx=\int u_0(x)\, dx$, and the contraction property in the $L_p(\mathbb R)$ space, 
\begin{equation}\label{Lp:es}
\|u(t)\|_p\leq \|u_0\|_p,
\end{equation}
for each $p\in [1,\infty]$, and all $t>0$. Moreover, the maximum and minimum principles hold, that is, 
\begin{equation}\label{29/12/2014-3.6}
\mbox{\rm ess}\,\inf u_0\le u(x,t) \le \mbox{\rm ess}\, \sup u_0,\ \ \mbox{a.e.}\ x,t,
\end{equation}
and the comparison principle is valid, which means that if  $u_0\le v_0\in L^1({\mathbb R} )$,  then 
\begin{equation}\label{comp}
u(x,t)\le v(x,t)\ \ \mbox{a.e.}\ \  x, \,t, \mbox{ and}\ \ \|u(t)-v(t)\|_1\le\|u_0-v_0\|_1.
\end{equation}

\noindent
(ii) Under the following additional conditions on the symbol of $\cal A$, 
 \begin{equation}\label{sym}
0<\liminf_{\omega\to 0}{{\psi(\omega)}\over{|\omega|^\alpha}}\le \limsup_{\omega\to 0}{{\psi(\omega)}\over{|\omega|^\alpha}}<\infty, \qquad
0<\inf_\omega{{\psi(\omega)}\over{|\omega|^2}},
\end{equation}
for some $0<\alpha<2$,  the more precise bound,   
\begin{equation*}
\|u(t)\|_p\le C_p\min(t^{- (1-1/p)/2},t^{- (1-1/p)/\alpha})\|u_0\|_1
\end{equation*}
holds for all $1\le p\le\infty$. Moreover, if $u_0\in L^1({\mathbb R} )\cap L^\infty({\mathbb R} )$, then
\begin{equation}\label{Lp:decay}
\|u(t)\|_p\le C(1+t)^{- (1-1/p)/\alpha}
\end{equation}
with a constant $C$ which depends only on $\|u_0\|_1$ and $\|u_0\|_p$.\\

\noindent
(iii) Assume that $u$ is a solution of the Cauchy problem \eqref{L1} with $u_0\in L^1({\mathbb R} )\cap L^\infty({\mathbb R} )$,  and that the symbol $\psi$ of the generator $\cal A$ satisfies Eqs. \eqref{infH} and \eqref{sym} with some $0 < \alpha < 2$. Furthermore, suppose that the nonlinearity $g$ is supercritical, that is, $g\in C^1$, and $\limsup_{s\to 0} |g(s)|/|s|^r$ $<\infty$,  for some $r > {\rm{max}}(\alpha ,1)$. Then the relation
\begin{equation}\label{first}
\lim_{t\to \infty}t^{ (1-1/p)/\alpha}\|u(t)-e^{-t{\cal A}}u_0\|_p = 0
\end{equation}
holds for every $1\le p\le\infty$. As usual,   $e^{-t{\cal A}}u_{0}$ denotes the action of the L\'evy semigroup on the function $u_0$, i.e. is a solution of the linear  {Eq. \eqref{29/12/2014-4}} with the initial data $u_{0}$. 
\end{thm}

On the other hand, the asymptotics of  the  solution of  the linear Cauchy problem {Eq. \eqref{29/12/2014-4}} is well known:  there exists a nonnegative function $\eta\in L^\infty(0,\infty)$ satisfying
$\lim\limits_{t\to\infty} \eta(t)=0$ such that
\begin{equation}
\Bigl \| e^{ {-}t{\cal A}}*u_0- \int_{{\mathbb R}} u_0(x)\,dx \cdot p_{\cal A}(t)\Bigr\|_p\le t^{-(1-1/p)/\alpha}\eta(t),
\end{equation}
where $p_{\cal A}(t)$ is the kernel of the operator $\cal A$ in Eq.  \eqref{29/12/2014-3} . Higher order asymptotics is also available \cite{BKW1999}.

The above general results have  direct consequences for {\it multifractal   conservation laws} driven by multiscale anomalous diffusions introduced in Section 2\footnote{The particle approximations and the propagation of chaos results for such systems have been studied in \cite{JMW2005}.}. Note the parabolic regularization included in the operator $\cal A$ because of the conditions \eqref{infH}, and \eqref{sym}.

\begin{cor}\label{c2.1}
All the statements of Theorem \ref{t3.1} are  valid for the conservation laws
\begin{equation}\label{LL1}
\frac{\partial u}{\partial t} + {\cal A}_ {(\vec \alpha;\vec \beta, \vec \gamma)}u + \frac{\partial}{\partial x} g(u) = 0, \qquad u(x,0) = u_0(x), 
\end{equation}
with 
\begin{equation*}
\alpha= \alpha_1<\alpha_2<\dots <\alpha_n  =2.
\end{equation*}
In particular, if  $u$ is a solution of the Cauchy problem \eqref{LL1} with $u_0\in L^1({\mathbb R} )\cap L^\infty({\mathbb R} )$,  and the nonlinearity $g \in C^1$ is supercritical, i.e., $\limsup_{s\to 0} |g(s)|/|s|^r$ $<\infty$, for $r > {\rm{max}}(\alpha ,1)$, then the relation
\begin{equation}\label{second}
\lim_{t\to \infty}t^{ (1-1/p)/\alpha}\|u(t)-e^{-t{\cal A}}u_0\|_p = 0
\end{equation}
holds for every $1\le p\le\infty$. Moreover, 
\begin{equation*}
\Bigl \| e^{t{\cal A}_ {(\vec \alpha;\vec \beta, \vec \gamma)}}*u_0- \int_{{\mathbb R}} u_0(x)\,dx \cdot p_{{\cal A}_ {(\vec \alpha;\vec \beta, \vec \gamma)}}(t)\Bigr\|_p\le t^{-(1-1/p)/\alpha}\eta(t),
\end{equation*}
where $p_{{\cal A}_ {(\vec \alpha;\vec \beta, \vec \gamma)}}(t)$ is the kernel of the operator ${\cal A}_ {(\vec \alpha;\vec \beta, \vec \gamma)}$ in Eq. \eqref{LL1}.   
\end{cor}

To prove Corollary \ref{c2.1}   it suffices to show that   conditions Eq. \eqref{infH} and Eq. \eqref{sym} are satisfied. Indeed, for the mutiscale  L\'evy process with   symbol \eqref{multisymbol}, we have
\begin{equation*}
\limsup_{|\omega|\to\infty} {\psi_ {(\vec \alpha;\vec \beta, \vec \gamma)}(\omega)-\gamma_n|\omega|^2 \over   {|\omega|^{ \alpha^*} }} = a_{j^*} < \infty
\end{equation*}
with $\alpha_{j^*}=\alpha^*$ where $  \alpha^*=\max (\alpha_1,\dots,\alpha_{(n-1)})$. Also, 
\begin{equation*}
0 < \lim_{\omega\to 0}{{ \psi_ {(\vec \alpha;\vec \beta, \vec \gamma)}(\omega)}\over{|\omega|^ {\alpha_*}}} = a_{j_*} <\infty,
\end{equation*}
with $\alpha_{j_{*}} = \alpha_{*}$ where $\alpha_* = \min (\alpha_1,\dots,\alpha_n) = \alpha_1$; and 
\begin{equation*}
 \inf_\omega{{ \psi_ {(\vec \alpha;\vec \beta, \vec \gamma)}(\omega)}\over{|\omega|^2}}\ge \gamma_n>0.  
\end{equation*}

\medskip

The above results depended on the subcritical behavior 
$$
 \limsup_{s\to 0}\frac{ |g(s)|}{|s|^r} <\infty , \qquad {\rm  for} \qquad r > {\rm{max}}(\alpha ,1), 
 $$
 of the nonlinearity in the conservation laws discussed above. Note that in the classical case of the Burgers equation the situation is dramatically different. 
 
 \bigskip

{\bf  Remark 1. \it Asymptotics of solutions of the Burgers equation.
{\rm    The first order asymptotics of solutions of the Cauchy problem for the Burgers equation  
\begin{equation}\label{burgers}
\frac{\partial}{\partial t} u(t, x) - \frac{\partial^{2}}{ \partial x^{2}}u(t, x) + \frac{\partial}{\partial x} [u(t, x)]^2 = 0
\end{equation}   
is described by the relation
\begin{equation*}
t^{(1-1/p)/2}\|u( t)-U_M( t)\|_p\to 0,\quad {\rm as} \quad t\to\infty,
\end{equation*}
where
\begin{equation*}
U_M(x,t) = \frac{e^{-x^2/(4t)}}{t^{1/2}} \left(K(M) + \frac{1}{2} \int_0^{\ulamek{x}{2} \sqrt t} e^{-\omega^2/4}\, d\omega\right)^{-1}
\end{equation*}
is the so-called source solution with the initial condition  $u(x, 0) = M\delta_0$. It is easy to verify that this solution is self-similar, i.e., $U_M(x,t)=t^{-1/2}U(xt^{-1/2},1)$. Thus, the long time behavior of solutions of Eq. \eqref{burgers} is genuinely nonlinear, i.e., it is not determined by the asymptotics of the linear heat equation. This strongly nonlinear behavior is due to the  precisely matched balancing influence of the regularizing Laplacian diffusion operator and the gradient-steepening quadratic inertial nonlinearity, see \cite{Z1993, WAW1998, WAW2001}.

\bigskip

Although not needed explicitly in the remainder of the paper, for the sake of completeness  we are providing below a general result  showing how such a matching  critical nonlinearity exponent for the nonlocal multifractal conservation law  yields  the solutions of (\ref{L1}) which behave asymptotically like  the self-similar source solutions $U$  of (\ref{L1}) with singular initial data $M\delta_0$.

\begin{thm}\label{t2} 
(see {\rm \cite{BKW2001a}}) Let $1 < \alpha < 2$, and $u$ be a solution of the Cauchy problem (\ref{L1}) with  the operator ${\cal A} = (-\Delta)^{\alpha/2} + {\cal K}$,  with the perturbation  ${\cal K}$ being  another L\'evy infinitesimal generator whose symbol $k$ fulfills the condition, 
\begin{equation}\label{critical}
\lim_{\omega\to 0}{k(\omega)\over |\omega|^\alpha} = 0,
\end{equation}
and $u_0\in L^1({\mathbb R} )$, $\int_{{\mathbb R}^d} u_0(x)\, dx = M > 0$. Assume that $g$ satisfies the condition
\begin{equation}
\lim_{s \to 0}{g(s)\over s|s|^{(\alpha-1) }}\in {\mathbb R}.
\end{equation}
Then, for each $1\le p\le\infty$,
\begin{equation}
\lim_{t\to\infty}t^{ (1-1/p)/\alpha}\|u(t)-U(t)\|_p= 0,
\end{equation}
where $U=U_M$ is the unique solution of the problem (\ref{L1}) with $r=\alpha$ and the initial data $M\delta_0$.  Moreover, $U$ is of self-similar form $U(x,t)=t^{-1/\alpha}U(xt^{-1/\alpha},1)$, $\int_{{\mathbb R}^d}U(x,1)\, dx= M$, and $U\ge 0$.
\end{thm}
Thus, analogous to Corollary \ref{c2.1}, we also have the following result in the   case of multifractal conservation laws with critical nonlinearities. Note  that, in contrast to Corollary \ref{c2.1}, the parabolic regularization is not necessary here.

\begin{cor}\label{2.2}
All the statements of Theorem \ref{t2} are  valid for the multifractal conservation laws 
 \begin{equation}
\frac{\partial u}{\partial t} + {\cal A}_ {(\vec \alpha;\vec \beta, \vec \gamma)}u + \frac{\partial}{\partial x}g(u) = 0, \qquad u(x,0) = u_0(x), \label{LL}
\end{equation}
with  $\alpha=\alpha_*\equiv \min(\alpha_1,\dots,\alpha_k)=\alpha_1$.
\end{cor}
 
 {
The verification of the condition (\ref{critical}) is immediate.  With the symbol of the perturbation $\cal K$, 
\begin{equation*}
{k}(\omega) =\sum_{j=2}^n -  \gamma_j|\omega|^{\alpha_{j}} e^{\frac{i\pi}{2} \beta_{j}\, {\rm sgn}(\omega)}
\end{equation*}
we do have $\lim\limits_{\omega\to 0}{ k(\omega) }/ |\omega|^\alpha=0$. Recall that,  in view of the convention adopted at the beginning of the paper,   $\alpha_*=\alpha_1<\alpha_2<\dots\alpha_n$.}

\medskip

 {The issue of explicit representations of source solutions of fractal conservation laws with critical nonlinearities is obviously  more difficult than the problems we are addressing in the subsequent sections, but  we plan to investigate it in the future.  }

 %--------------------------------------------------    Sec 4   --------------------------------------------------------
 
\section{Explicit representation of the kernels of  the two-scale, two-sided L\'evy generators, $0 < \alpha \leq 2$}

In this section our goal is to find   explicit representations for kernels $ {v}_ {(\vec \alpha;\vec \beta, \vec \gamma)}$ of the infinitesimal generators   ${\cal A}_ {(\vec \alpha;\vec \beta, \vec \gamma)}$  which dictate the long-time behavior of the nonlinear conservation laws discussed in Section 3. For the sake of simplicity, we present the case when the scaling parameter $\vec \gamma =(1,\dots,1)$; the notation is then streamlined   to   $ {v}_ {(\vec \alpha;\vec \beta, \vec \gamma)}\equiv  {v}_ {(\vec \alpha;\vec \beta )}$. Simply stated, we need to find an explicit expression  for  the Fourier convolution of $v_{\alpha_{j}, \beta_{j}}(t, x)$, $j=1, 2$, $x\in\mathbb{R}$, and $t > 0$:
\begin{equation}\label{30Jun14-3}
H(t, x) = \int_{-\infty}^{\infty} v_{\alpha_{1}, \beta_{1}}(t, y) v_{\alpha_{2}, \beta_{2}}(t, x-y) dy = \int_{-\infty}^{\infty} v_{\alpha_{1}, \beta_{1}}(t, x-y) v_{\alpha_{2}, \beta_{2}}(t, y) dy,
\end{equation}
where $H(t, x) = H(\vec \alpha;\vec \beta; t, x)$,   $\vec \alpha=(\alpha_1,\alpha_2)$, and $\vec \beta=(\beta_1,\beta_2)$. The basic properties of $v_{\alpha_{j}, \beta_{j}}(t, x)$, with necessary conditions on $\alpha_{j}$ and $\beta_{j}$, are given in Section 2.

The functions $v_{\alpha_{j}, \beta_{j}}(t, x)$, $j=1, 2$, represent the unimodal probability density functions of two-sided L\'{e}vy stable distributions \cite{ELukacs70, WFeller},  which correspond to one-sided L\'evy stable distributions for $0 < \alpha_{j} < 1$ and $\beta_{j} = -\alpha_{j}$. This case will be discussed in section 5. The series representation of two-sided L\'{e}vy stable distributions for $0 < \alpha_{j} < 1$, and $|\beta_{j}| \leq \alpha_{j}$, can be found  in, e.g., Eq. (5.8.8a) on p. 142  \cite{ELukacs70}, Eq. (6.8) on p. 583 of \cite{WFeller}, and Eq. (4) in  \cite{HBerstrom52}, whereas, for $ 1 < \alpha_{j} \leq 2$, and $|\beta_{j}| \leq 2-\alpha_{j}$,  they are described in, e.g., Eq. (5.8.8b) on p. 142 of \cite{ELukacs70},  and Eq. (6.9) on p. 583 of \cite{WFeller}. Those two different types of series expansions were calculated for rational values of parameter $\alpha_{j}$ and $\beta_{j}$, see Eqs. (4) and (5) in \cite{KGorska11}. We quote some solution which will be used later in the paper: the Gaussian distribution 
\begin{equation}\label{29/12/2014-a}
v_{2, 0}(t, x) = \frac{\exp(-\ulamek{x^2}{4t})}{2\sqrt{\pi t}}
\end{equation}
for $\alpha = 2$ and $\beta=0$, the L\'{e}vy-Smirnov distribution
\begin{align}\label{26/09/2014-1a}
v_{\frac{1}{2}, -\frac{1}{2}}(t, x) & = \frac{t \exp(-\ulamek{t^2}{4x})}{2\sqrt{\pi} x^{3/2}} , \qquad x>0 \\ & = 0,   \qquad\qquad\qquad\,\, x \leq 0.
\end{align}
for $\alpha = 1/2$ and $\beta = -1/2$, and 
\begin{align}\label{29/12/2014-b}
v_{\frac{3}{2}, -\frac{1}{2}}(t, x) & = \frac{(2/t)^{2/3}}{3\sqrt{\pi}} \frac{\Gamma(\ulamek{5}{6})}{\Gamma(\ulamek{2}{3})}\, {_{1}F_{1}}\left({5/6 \atop 2/3}; -\frac{4 x^3}{27 t^2}\right) + \frac{(2/t)^{4/3}}{9\sqrt{\pi}} \frac{\Gamma(\ulamek{7}{6})}{\Gamma(\ulamek{4}{3})}\, x\, {_{1}F_{1}}\left({7/6 \atop 4/3}; -\frac{4 x^3}{27 t^2}\right) \nonumber\\
& = \frac{Re}{2\pi} \int_{-\infty}^{\infty} e^{-i \omega x} e^{-t|\omega|^{3/2}\exp(-\frac{i\pi}{4} {\rm sgn}(\omega)} d\omega 
\end{align}
for $\alpha = 3/2$ and $\beta = -1/2$ \cite{KGorska11}.  The symbol ${_{1}F_{1}}$ stands for  the hypergeometric function introduced in Section 1.

Let us now find the explicit form of $H(t, x)$ given in Eq. \eqref{30Jun14-3}. Applying the property \eqref{29/12/2014-8a} to Eq. \eqref{30Jun14-3}, we can rewrite $H$ in the form, 
\begin{align}\label{30Jun14-4}
H(t, x)  = \mathcal{F}^{-1}[\tilde{v}_{\alpha_{1}, \beta_{1}}(t, \omega)\, \tilde{v}_{\alpha_{2}, \beta_{2}}(t, \omega); x] = H_{-}(t, -x) \Theta(-x) + H_{+}(t, x)\Theta(x),
\end{align}
where 
\begin{equation}\label{30/06/14-5}
H_{+}(t, x)  = H_{+}({\alpha_{1}, \beta_{1}, \alpha_{2}, \beta_{2}}; t, x) = \frac{\textit{Re}}{\pi} \int_{0}^{\infty} \hskip -2mm e^{-i x \omega} \exp\left(\!- t \omega^{\alpha_{1}} e^{\frac{i\pi}{2}\beta_{1}} - t \omega^{\alpha_{2}} e^{\frac{i\pi}{2}\beta_{2}}\right) d\omega, 
\end{equation}
and 
\begin{equation}\label{30/06/14-5a}
H_{-}(t, x)  = H_{+}(\alpha_{1}, -\beta_{1}, \alpha_{2}, -\beta_{2}; t, x).
\end{equation}
The function $\Theta(x)$ is here the usual Heaviside step function. The matching, at $x=0$,  of these two components is assured by the continuity at the origin of $H(t, x)$, and of all of its higher derivatives. Indeed, the continuity of Eq. \eqref{30Jun14-4} at $x=0$ can be shown by employing  Eq. \eqref{30/06/14-5} as follows: for $n=0, 1, 2, \ldots$
\begin{align}\label{30/07/14-3}
\partial^{n}_{x} H_{+}(t, x)\big\vert_{x=0} & = \frac{\textit{Re}}{\pi} \int_{0}^{\infty} (-i \omega)^{n} e^{- t \omega^{\alpha_{1}}\exp\big(\ulamek{i\pi}{2} \beta_{1}\big) - t \omega^{\alpha_{2}}\exp\big(\ulamek{i\pi}{2} \beta_{2}\big)} d\omega \nonumber \\
& = \frac{1}{\pi} \int_{0}^{\infty} e^{- t \omega^{\alpha_{1}} \cos\!\big[\ulamek{\pi}{2}(\beta_{1} - n)\big] - t \omega^{\alpha_{2}} \cos\!\big[\ulamek{\pi}{2}(\beta_{2} - n)\big]}  \nonumber \\
& \times \cos\left(t \omega^{\alpha_{1}} \sin\big[\ulamek{\pi}{2}(\beta_{1} - n)] + t \omega^{\alpha_{2}} \sin\big[\ulamek{\pi}{2}(\beta_{2} - n)]\right) d\omega \nonumber \\
& = \frac{\textit{Re}}{\pi} \int_{0}^{\infty} (i \omega)^{n} e^{- t \omega^{\alpha_{1}}\exp\big(-\ulamek{i\pi}{2} \beta_{1}\big) - t \omega^{\alpha_{2}}\exp\big(-\ulamek{i\pi}{2} \beta_{2}\big)} d\omega \nonumber \\
& = \partial^{n}_{x} H_{-}(t, -x)\big\vert_{x=0}.
\end{align}
We would also like to point out that Eqs. \eqref{30Jun14-4} and \eqref{30/06/14-5},  in the case  $\alpha_{1} = \alpha_{2}$, and $\beta_{1} = \beta_{2}$, imply the identity, 
\begin{equation}
\label{30/07/14-5}
H(x, t) = v_{\alpha_{1}, \beta_{1}}(x, 2t).
\end{equation}

In what follows, without loss of  generality, we will consider only the case of $H_{+}(x, t)$. The function $H_{-}(x, t)$ will be used only when necessary. We assume that, for certain values of complex $s$, the Mellin transform of $H_{+}(x, t)$ exists and, according to the notation introduced in Eq. \eqref{26/08/14-3a}, it is denoted by $\hat{H}_{+}(s, t)$. Thereafter, we substitute Eq. \eqref{30/06/14-5} into $\hat{H}_{+}(s, t)$, change the order of integration and use the one of Eqs. (2.3.2.13)  of \cite{APPrudnikov-v1}. Those steps imply that for rational $\alpha_{j}$ and $\beta_{j}$, $j=1, 2$, such that $\alpha_{1} = \ulamek{l}{k}$, $\beta_{1} = \ulamek{l-2a}{k}$, $\alpha_{2} = \ulamek{p}{q}$, and $\beta_{2} = \ulamek{p-2b}{q}$, where $l$, $k$, $p$, $q$, $a$, and $b$, are integers,  we have
\begin{align}
\label{30/06/14-6a}
\hat{H}_{+}(s, t) & = \frac{1}{M \pi}\sum_{j=0}^{M_{1}-1} \frac{(-1)^{j}}{j!}\, \frac{\Gamma(s)\, \Gamma\big(\frac{1-s}{M} + \frac{m}{M}j\big)}{t^{\frac{1-s}{M} + (\frac{m}{M}-1)j}}\, {\textit Re}\left\{e^{-i\pi\big[\frac{1}{2} - u\frac{1-s}{M} + \big(v-u\frac{m}{M}\big)j\big]} \right.
\nonumber\\
&\qquad  \left.\times {_{1+m_{1}}F_{M_{1}}}\left({1, \Delta(m_{1}, \frac{1-s}{M} + \frac{m}{M}j) \atop \Delta(M_{1}, 1+j)}; \big(\!-\ulamek{t e^{-i\pi v}}{M_{1}}\big)^{M_{1}} \big(\ulamek{m_{1}}{t e^{-i\pi u}}\big)^{m_{1}}\right)\right\}
 \\
& = \frac{1}{M \pi} \sum_{j=0}^{M_{1}-1}\sum_{r=0}^{\infty} \frac{(-1)^{j+r M_{1}}}{(j + r M_{1})!}\, \frac{\Gamma(s)\, \Gamma[\frac{1-s}{M} + \frac{m}{M}(j+r M_{1})]}{t^{\frac{1-s}{M} + (\frac{m}{M}-1)(j+r M_{1})}} \nonumber
 \\
&\qquad \times \sin\left[\pi u\ulamek{1-s}{M} - \pi (j+rM_{1})\big(v - u\ulamek{m}{M}\big) \right]
 \label{30/06/14-6b} 
\\ 
& = \frac{1}{M \pi} \sum_{r=0}^{\infty} \frac{(-1)^{r}}{r!} \, \frac{t^{-\frac{1-s}{M} - r(\frac{m}{M}-1)}\, \Gamma(s)\, \Gamma(\ulamek{1-s}{M} + \ulamek{m}{M}r)}{\Gamma\big[1- u \ulamek{1-s}{M} + (v-u\ulamek{m}{M})r\big]\, \Gamma\big[u \ulamek{1-s}{M} - (v-u\ulamek{m}{M})r\big]}, 
\label{30/06/14-6c}
%\cos\left[\pi\big(\ulamek{1}{2} - u\ulamek{1-s}{M}\big) + \pi r \big(v - u\ulamek{m}{M}\big) \right], 
\end{align}
where $m$, $M$, $m_{1}$, and $M_{1}$ are  as follows:
\begin{equation}
\label{26/08/14-10}
m = \min(\ulamek{l}{k}, \ulamek{p}{q}), \quad M = \max(\ulamek{l}{k}, \ulamek{p}{q}), \quad m_{1} = \min(kp, lq), \quad \text{and} \quad M_{1} = \max(kp, lq).
\end{equation}
The parameters $u$, and $v$, are determined by the equalities,   
\begin{equation}
\label{31/08/14-1}
u = \frac{a}{k}, \,\, v = \frac{b}{q}, \,\,\,  {\rm for} \,\,\, \alpha_{1} > \alpha_{2}, \quad {\rm and} \quad u = \frac{b}{q}, \,\, v = \frac{a}{k} \,\,\,  {\rm for} \,\,\, \alpha_{1} < \alpha_{2}.
\end{equation}
In Eq. \eqref{30/06/14-6a} we utilized a series representation of the generalized hypergeometric function given in  Eq. (\ref{26/08/14-6}),  and the Gauss-Legendre multiplication formula defined in Eq. (\ref{26/08/14-8}). In Eq. \eqref{30/06/14-6b} we applied Eq. \eqref{26/08/14-7},  and also changed the summation index as follows: $j+r M_{1} \to r$.

The next step requires inverting the Mellin transform in Eq. \eqref{30/06/14-6c}. To accomplish this  task  we will introduce   the new variable of integration,  $\tilde{s} = (1-s)/(lp)$, Eqs. \eqref{26/08/14-7}, and \eqref{26/08/14-8}. Putting the all of these terms together, we get,  for $x > 0$, 
\begin{align}
\label{30/06/14-7}
H_{+}(x, t) & =  \frac{m_{1}\sqrt{M}}{x (2\pi)^{\frac{lp+m_{1}}{2} - u m_{1}}} \sum_{r=0}^{\infty} \frac{(-t)^{r}}{r!} \left(\frac{m_{1}}{t}\right)^{\!\frac{m}{M}r}\nonumber 
\\&\qquad  \times G^{m_{1}, lp}_{lp + u m_{1}, m_{1} + u m_{1}}\left(\frac{(lp)^{lp}\, t^{m_{1}}}{x^{lp}\, m_{1}^{m_{1}}} \Big\vert {\Delta(lp, 0), \Delta(u m_{1}, -(v-u\frac{m}{M} r)) \atop \Delta\big(m_{1}, \frac{m}{M}r\big), \Delta(u m_{1}, -(v-u\frac{m}{M} r))}\!\right).
\end{align}
The Meijer $G$ functions in Eq. \eqref{30/06/14-7} can be expressed, via formulas (8.2.2.3), and (8.2.2.4), on p. 520 of \cite{APPrudnikov-v3}, in terms of a generalized hypergeometric function. With respect to the values of $\alpha_{1}$ and $\alpha_{2}$, we can consider two different cases:

\begin{itemize}
\item[\textbf{(A)}] 
After applying Eq. (8.2.2.3) on p. 520 of  \cite{APPrudnikov-v3} to Eq.~\eqref{30/06/14-7},  for $0 < \alpha_{i} < 1$, $i = 1, 2$, we have
\begin{align}
\label{30/07/14-1}
H_{+}(x, t) &= -\frac{1}{\pi} \sum_{r=0}^{\infty} \sum_{j=0}^{m_{1}-1} \frac{(-t)^{r+j}}{r!\, j!} \frac{\Gamma(1 + Mj + mr)}{x^{1 + Mj + rm}} \sin(r\pi v + j\pi u) \nonumber 
\\
& \quad \times {_{1+lp} F_{m_{1}}}\left({1, \Delta(lp, 1+ Mj + mr) \atop \Delta(m_{1}, 1 + j)}; (-1)^{m_{1}u - m_{1}} \frac{t^{m_{1}} (lp)^{lp}}{m_{1}^{m_{1}} x^{lp}}\right),
\end{align}
where $u$, and $v$, are given in Eq. \eqref{31/08/14-1}\footnote{Let us observe that, for $u = M$, and $v = m$, and thus, for  $a = l$, and $b = p$, Eq. \eqref{30/07/14-1} gives Eq. \eqref{14/07/14-2} of Section 5, this is the Laplace convolution of two one-sided L\'{e}vy stable distributions.}. Moreover, using the series expansion  of  the function ${_{1+lp} F_{m_{1}}}$, Eq. \eqref{30/07/14-1}, can be expressed as follows: 
\begin{equation}
\label{30/07/14-2}
H_{+}(x, t) = -\frac{1}{\pi} \sum_{r, n = 0}^{\infty} \frac{(-t)^{n + r}}{n!\, r!}\, \frac{\Gamma(1 + \alpha_{1} r + \alpha_{2} n)}{x^{1 + \alpha_{1} r + \alpha_{2} n}}\, \sin\big(\pi r\,  \ulamek{\alpha_{1} - \beta_{1}}{2} + \pi n\,  \ulamek{\alpha_{2} - \beta_{2}}{2}\big),
\end{equation}
which for $t=1$, $x > 0$, and $r=0$ (or $n=0$), is identical with the series expression for two-sided L\'{e}vy stable distribution given in, e.g.  Eq. (5.8.8a) on p. 142, in \cite{ELukacs70}.

\item[\textbf{(B)}] 
For $ 1 < \alpha_{i} \leq 2$, $i = 1, 2$, Eq. (8.2.2.4) on p. 520 of \cite{APPrudnikov-v3},  applied to  Eq. \eqref{30/06/14-7} gives
\begin{align}
\label{30/07/14-4}
H_{+}(x, t) &= \frac{1}{\pi M} \sum_{r=0}^{\infty} \sum_{j=0}^{lp -1} \frac{(-1)^{r+j}}{r!\, j!} \frac{x^{j}}{t^{\frac{1+j}{M} + (\frac{m}{M} - 1)r}}\, \Gamma\big(\ulamek{1+j}{M} + \ulamek{m}{M}r\big) \sin\big[\pi u\ulamek{1+j}{M} - \pi r \big(v - u \ulamek{m}{M}\big)\big] \nonumber \\[0.2\baselineskip]
& \quad \times {_{1+m_{1}}F_{lp}}\left({1, \Delta(m_{1}, \ulamek{1+j}{M} + \ulamek{m}{M} r) \atop \Delta(lp, 1+j)}; (-1)^{m_{1}u + lp} \frac{m_{1}^{m_{1}} x^{lp}}{t^{m_{1}} (lp)^{lp}}\right),
\end{align}
which can be rewritten as
\begin{equation}
\label{30/07/14-5}
H_{+}(x, t) = \frac{1}{\pi M} \sum_{r, n = 0}^{\infty} \frac{(-1)^{r+n}}{r!\, n!} \frac{x^{n}}{t^{\frac{1+n}{M} + (\frac{m}{M} - 1)r}}\, \Gamma\big(\ulamek{1+n}{M} + \ulamek{m}{M}r\big)\,\sin\big[\pi u\ulamek{1+n}{M} - \pi r \big(v - u \ulamek{m}{M}\big)\big].
\end{equation} 
\end{itemize}

\noindent
{\it Example 4.1. The bi-Gaussian case.} 
The  elementary case $\alpha_{1} = \alpha_{2} = 2$, $\beta_{1} = \beta_{2} = 0$,  is straightforward and we include it here only for verification's sake. Substituting Eq. \eqref{29/12/2014-a} into Eq. \eqref{30Jun14-3} and employing Eq. (3.323.2) on p. 337 of \cite{Gradshteyn}, we get
\begin{align} \label{30/12/2014-aa}
%\int_{-\infty}^{\infty} v_{2, 0}(y, t) v_{2, 0}(x-y, t) dy 
H(2, 0, 2, 0; t, x) = \frac{\exp(-\frac{x^{2}}{4t})}{4\pi t} \int_{-\infty}^{\infty} e^{-\ulamek{y^{2}}{2 t} + \ulamek{x y}{2t}} dy = \frac{\exp(-\ulamek{x^{2}}{8 t})}{2 \sqrt{2\pi t}}  = v_{2, 0}(x, 2t),
\end{align}
which is in agreement with Eq. \eqref{30/07/14-5} and is presented in Fig. \ref{fig2}, see the curve I (red).\\

\noindent
{\it Example 4.2. The Gaussian-L\'{e}vy case.} Here  $\alpha_{1} =   2$, $\alpha_{2} =   1/2$, $\beta_{1} = 0$,  and $\beta_{2} = -1/2$,  so that Eq. \eqref{30Jun14-3} reads
  \begin{align}
%& \int_{-\infty}^{\infty} v_{\frac{1}{2}, -\frac{1}{2}}(y, t) v_{2, 0}(x-y, t) dy 
H(2, 0, \ulamek{1}{2}, -\ulamek{1}{2}; t, x) & = \frac{1}{4 \pi \sqrt{t}} e^{-\frac{x^2}{4 t}} \int_{0}^{\infty} y^{-\frac{3}{2}} e^{-\frac{t^2}{4 y}} e^{-\frac{y^2}{4 t} + \frac{x y}{2 t}} dy \nonumber \\
& = \frac{\sqrt{t}}{4\pi} e^{-\frac{x^2}{4 t}} \sum_{r=0}^{\infty} \frac{(-t^2/4)^r}{r!} \int_{0}^{\infty} y^{-\frac{3}{2}-r} e^{-\frac{y^2}{4 t} + \frac{x y}{2 t}} dy \label{10/12/2014a} \\
& = \frac{1}{2} e^{-\frac{x^2}{8 t}} \sum_{r=0}^{\infty} \frac{(-1)^{2r+1}}{r! \Gamma(\frac{3}{2} + r)}\, \frac{t^{\frac{1}{4} + \frac{3}{4}r}}{2^{\frac{5}{4} + \frac{5}{2}r}}\, D_{\frac{1}{2} + r}(-\ulamek{x}{\sqrt{2 t}}) \label{10/12/2014b} \\
& = \frac{1}{2\sqrt{\pi t}} e^{-\frac{x^2}{8 t}} \sum_{r=0}^{\infty} \frac{(-1)^{2r+1}}{(2r+1)!} \left(\!\frac{t^3}{2}\!\right)^{\!\frac{1}{4}(2r + 1)}\!\! D_{\frac{2r+1}{2}}(-\ulamek{x}{\sqrt{2 t}}) \label{10/12/2014c} \\
& = \frac{1}{2\sqrt{\pi t}} e^{-\frac{x^2}{8 t}} \sum_{n=0}^{\infty} \frac{(-1)^{n}}{n!} \left(\!\frac{t^3}{2}\!\right)^{\!\frac{n}{4}}\! D_{\frac{n}{2}}(-\ulamek{x}{\sqrt{2 t}}), \label{10/12/2014d}
\end{align}
where $D_{\nu}(z)$ is the parabolic cylinder function \cite{Gradshteyn}. In Eq. \eqref{10/12/2014a} we applied Eq. (9.241.2) on p. 1028 of \cite{Gradshteyn}, and in Eq. \eqref{10/12/2014c} we changed the summation index as follows:  $n = 2r + 1$. Eq. \eqref{10/12/2014d} can be obtained from Eq. \eqref{30/07/14-4} after using Eqs. (7.11.3.3) and (7.11.3.4) on p. 491 of \cite{APPrudnikov-v3}. The plot of $H(2, 0, \ulamek{1}{2}, -\ulamek{1}{2}; t, x)$ for $t=1$ is illustrated in Fig. \ref{fig2}, see the curve II (blue).
\\

\noindent
{\it Example 4.3.} In Eq. \eqref{30Jun14-3} we take $v_{\alpha_{j}, \beta_{j}}(t, x)$, $j=1, 2$, given in Eqs. \eqref{29/12/2014-a} and \eqref{29/12/2014-b}. Thus, we get
\begin{align}\label{30/12/2014-1a} 
& H(2, 0, \ulamek{3}{2}, -\ulamek{1}{2}; t, x) = \int_{-\infty}^{\infty} \frac{\exp(-\frac{y^2}{4 t})}{2\sqrt{\pi t}} \frac{Re}{2\pi} \left\{ \int_{-\infty}^{\infty} e^{-i \omega (x-y)} e^{-t|\omega|^{3/2}\exp[-\frac{i\pi}{4} {\rm sgn}(\omega)]} d\omega \right\} dy \\ \label{30/12/2014-1b}
& \qquad = \frac{Re}{2\pi} \int_{-\infty}^{\infty} e^{-i \omega x} e^{-t|\omega|^{3/2}\exp[-\frac{i\pi}{4} {\rm sgn}(\omega)]} \left[\int_{-\infty}^{\infty} e^{i\omega y}\frac{\exp(-\frac{y^2}{4 t})}{2\sqrt{\pi t}} dy\right] d\omega \\ \label{30/12/2014-1c}
& \qquad = \frac{Re}{2\pi} \int_{-\infty}^{\infty} e^{-i \omega x} e^{-t|\omega|^{3/2}\exp[-\frac{i\pi}{4} {\rm sgn}(\omega)] - t \omega^2} d\omega,
\end{align}
which is in agreement with Eq. \eqref{30Jun14-4} for $\alpha_{1} = 2$, $\alpha_{2} = 3/2$, $\beta_{1} = 0$, and $\beta_{2} = -1/2$. Thus we can repeat all the steps from Eq. \eqref{30Jun14-4} to Eq. \eqref{30/07/14-4} which  gives
\begin{align}\label{30/12/2014-2a}
H(2, 0, \ulamek{3}{2}, -\ulamek{1}{2}; t, x) & = \frac{1}{2\pi} \sum_{r=0}^{\infty} \sum_{j=0}^{5} \frac{(-1)^{r+j}}{r! j!} \frac{x^j}{t^{\frac{1+j}{2} - \frac{r}{4}}} \Gamma(\ulamek{1+j}{2} + \ulamek{3}{4} r) \cos(\ulamek{\pi}{2}j - \ulamek{\pi}{4}r) \nonumber \\
& \times {_{4}F_{6}}\left({1, \Delta(3, \ulamek{1+j}{2} + \ulamek{3}{4}r) \atop \Delta(6, 1+j)}; -\frac{x^6}{1728 t^3}\right). 
\end{align}
For $t=1$,  the function is plotted as  the curve III (green) in Fig. \ref{fig2}.

\begin{figure}[!h]
\begin{center}
\includegraphics[scale=0.6]{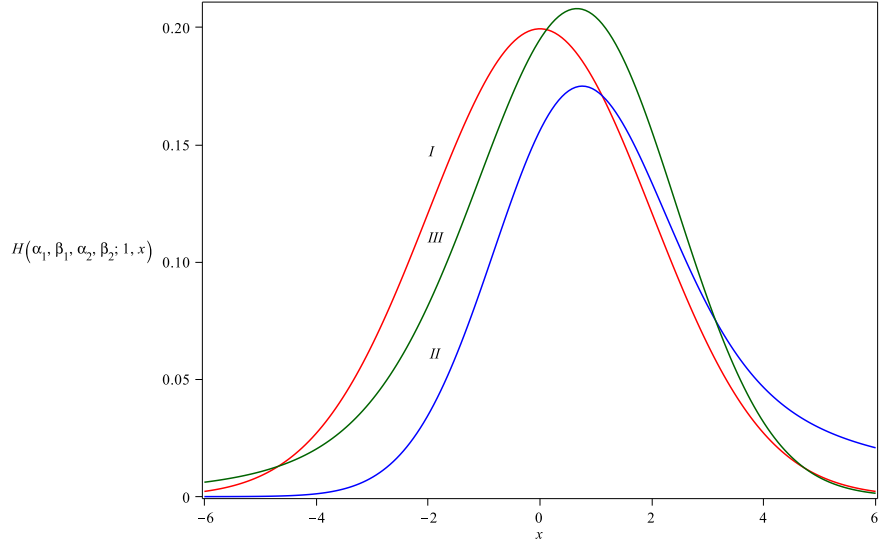}
\caption{\label{fig2} Multiscale densities $H(\alpha_{1}, \beta_{1}, \alpha_{2}, \beta_{2}; t, x)$, for $t = 1$, and given values of $\alpha_{1}$, $\beta_{1}$, $\alpha_{2}$, and $\beta_{2}$. In plot I (red),   $\alpha_{1} = \alpha_{2} = 2$, and $\beta_{1} = \beta_{2} = 0$, see Eq. \eqref{30/12/2014-aa}; in plot II (blue),  $\alpha_{1} = 2$, $\beta_{1} = 0$, $\alpha_{2} = \ulamek{1}{2}$, and $\beta_{2} = -\ulamek{1}{2}$, see Eq. \eqref{10/12/2014d}; in plot III (green), $\alpha_{1} = 2$, $\beta_{1} = 0$, $\alpha_{2} = \ulamek{3}{2}$, and $\beta_{2} = -\ulamek{1}{2}$, see Eq. \eqref{30/12/2014-2a}.
}
\end{center}
\end{figure}

%---------------------------------------------    Sec 5    ---------------------------------------------

\section{Explicit representation of the kernels of  the two-scale, one-sided L\'evy generators, $0<\alpha < 1$}

Although this case has no direct applicability to our work on the asymptotics of the multifractal conservation laws in Section 3, in anticipation of our future  work, in this section we provide explicit representations of the kernels of  the two-scale, one-sided L\'evy generators with $0<\alpha_1<\alpha_2\le 1$.  Here the tool is, of course, the Laplace transform. 

To further simplify our exposition  we will only consider two   one-parameter families,  $v_{\alpha_1}(t, x) \equiv v_{(\alpha_1, -\alpha_{1}, 1)}(t, x)$, and $v_{\alpha_2}(t, x) \equiv v_{(\alpha_2, -\alpha_{2}, 1)}(t, x)$,  of one-sided stable L\'evy densities whose the Laplace convolution has the form
\begin{equation}\label{25.06.14-1}
h_{\alpha_{1}, \alpha_{2}}(t, x) = \int_{0}^{x} v_{\alpha_{1}}(t, y) v_{\alpha_{2}}(t, x-y) dy =  \int_{0}^{x} v_{\alpha_{1}}(t, x-y) v_{\alpha_{2}}(t, y) dy.
\end{equation}
Functions $v_{\alpha_j}(t, x)$, $j =1, 2$ are given by the ``stretched exponential'' Laplace transform $\exp(- t p^{\alpha_{j}}) = \mathcal{L}[v_{\alpha_{j}}(t, x); p]$, see  \cite{RSAnderssen04, GDattoli14, HPollard46, KAPenson10}. Thus, the characteristic function  of  Eq.  \eqref{25.06.14-1} is of the form
\begin{equation}
\label{25.06.14-2}
\exp\left(-t p^{\alpha_{1}} - t p^{\alpha_{2}}\right) = \mathcal{L}[h_{\alpha_{1}, \alpha_{2}}(x, t); p].
\end{equation}
Eq. \eqref{25.06.14-2} implies that, for $0 < \alpha_{1} = \alpha_{2} < 1$, 
\begin{equation}
\label{26/08/14-9}
h_{\alpha_{1}, \alpha_{1}}(t, x) = v_{\alpha_{1}}(2t, x),
\end{equation} 
whereas $h_{\alpha_{1}, \alpha_{2}}(t, x)$ for arbitrary $\alpha_{j}$ can be found by using the series representation of a  one sided-L\'{e}vy stable distribution, 
\begin{equation}
\label{31/07/14-1}
v_{\alpha_{j}}(t, x) = -\frac{1}{\pi} \sum_{r=0}^{\infty} \frac{(-t)^{r}}{r!} \frac{\Gamma(1 + \alpha_{j} r)}{x^{1 + \alpha_{j} r}} \sin(\pi r \alpha_{j}).
\end{equation}
Eq. \eqref{31/07/14-1} follows from Eq. \eqref{31/08/14-1} applied for the one-sided L\'{e}vy stable distribution, and the formula  (4) of \cite{HPollard46}.

\bigskip

{\bf Remark 2. \it 
Absolute convergence of $v_{\alpha_{j}}(t, x)$.}  {\rm For $x \geq 0$,  the series in Eq. \eqref{31/07/14-1} converges absolutely and its radius of convergence is infinite. Indeed, the absolute value  of $v_{\alpha_{j}}(x, t)$, for $x > 0$,  can be estimated as follows:
\begin{equation*}
|v_{\alpha_{j}}(t, x,)| < \frac{1}{x} \sum_{n=0}^\infty \frac{\Gamma(1+\alpha_{j} n)}{n!}\! \left(\!\frac{t}{x^{\alpha_{j}}}\!\right)^{\!n} < \infty. 
\end{equation*}   
Next, from the Cauchy ratio test of convergence, \cite{GBArfken05},  and the Stirling formula, see Eq. (8.327.1) on p. 895 of \cite{Gradshteyn},   the absolute convergence of the series \eqref{31/07/14-1}  follows. The convergence of the series  \eqref{31/07/14-1}  at $x=0$ can be verified  by employing the asymptotic form of $v_{\alpha_{j}}(x, t)$ at $x=0$: 
\begin{equation}\label{28/08/14-2}
v_{\alpha_{j}}(t, x) \approx K t^{\ulamek{1}{2(1-\alpha_{j})}} x^{-\ulamek{2 - \alpha_{j}}{2 - 2\alpha_{j}}} \exp\!\Big(\!-A t^{\ulamek{1}{1 - \alpha_{j}}} x^{- \ulamek{\alpha_{j}}{1-\alpha_{j}}}\Big),
\end{equation}
where $K$ and $A$ are positive constants. Eq. \eqref{28/08/14-2} was obtained  with the help of Eq. (4) in \cite{JMikusinski59}. The absolute value of the right-hand side of  of Eq. \eqref{28/08/14-2}  obviously converges to 0, as $x\to 0+$ (for any fixed   $t > 0$, and $0 < \alpha_{j} < 1$).

\bigskip
 For rational $\alpha_{j} = l/k$, where $k$ and $l$ are positive integers,  Eq. \eqref{31/07/14-1} can be expressed via a finite sum of the generalized hypergeometric function, see Eqs. (3), and (4), in \cite{KAPenson10},  for $t=1$. Moreover, it turns out that, for $k \leq 3$,   it can be written down in terms of standard special functions, e.g., for $\alpha = 1/2$, 
\begin{equation}
\label{26/09/2014-1a}
v_{\frac{1}{2}}(t, x) = v_{\frac{1}{2}, -\frac{1}{2}}(t, x) \quad {\rm for}\,\, x > 0, 
\end{equation}
for $\alpha = 1/3$ \cite{HSher75} 
\begin{equation}\label{26/09/2014-1b}
v_{\frac{1}{3}}(t, x) = \frac{t^{3/2}}{3\pi x^{3/2}} K_{\frac{1}{3}}\!\left(\!\frac{2}{\sqrt{x}}\frac{t^{3/2}}{3^{3/2}}\right), 
\end{equation}
and for $\alpha = 2/3$ \cite{EWMontroll84}
\begin{equation}\label{26/09/2014-1c}
v_{\frac{2}{3}}(x, t) = \sqrt{\frac{3}{\pi}}\, \frac{\exp\big(\!-\ulamek{2 t^3}{27 x^2}\big)}{x} W_{\frac{1}{2}, \frac{1}{6}}\left(\frac{4 t^3}{27 x^2}\right),
\end{equation}
where $K_{\nu}(u)$ is a modified Bessel function of the second kind \cite{Gradshteyn}, and $W_{\nu, \mu}(u)$ is the Whittaker W function \cite{Gradshteyn}.

The substitution of Eq. \eqref{31/07/14-1} into Eq. \eqref{25.06.14-1} allows us to write, with help from   Eq. \eqref{26/08/14-7}, that 
\begin{align}
\label{8/07/14-1}
h_{\alpha_{1}, \alpha_{1}}(t, x) &= \int_{0}^{x} \sum_{r, j = 0}^{\infty} \frac{(-t)^{r+j}}{r! j!} \frac{\Gamma(1+\alpha_{1}j) \Gamma(1+\alpha_{2}r)}{y^{1 + j\alpha_{1}} (x-y)^{1 + r \alpha_{2}}} \sin(\pi\alpha_{1} j) \sin(\pi\alpha_{2} r) \frac{dy}{\pi^{2}} \nonumber \\
& = \sum_{r, j = 0}^{\infty} \frac{(-t)^{r+j}}{r! j!} \frac{1}{\Gamma(-\alpha_{1} j) \Gamma(-\alpha_{2} r)} \int_{0}^{x} y^{-1-j\alpha_{1}} (x-y)^{-1-r \alpha_{2}} dy \nonumber \\
& = -\frac{1}{\pi} \sum_{r, j = 0}^{\infty} \frac{(-t)^{r+j}}{r! j!}\, \frac{\Gamma(1+ \alpha_{1} j + \alpha_{2} r)}{x^{1+ \alpha_{1} j + \alpha_{2} r}}\, \sin(\pi\alpha_{1} j + \pi\alpha_{2} r).
\end{align}
Observe that  the expressions in Eq. \eqref{8/07/14-1}  are invariant with respect to the change of order of the parameters $\alpha_{1}$, and $\alpha_{2}$, so that $h_{\alpha_{1}, \alpha_{2}}(x, t) = h_{\alpha_{2}, \alpha_{1}}(x, t)$. Moreover, the first term in the sum in  Eq. \eqref{8/07/14-1}, corresponding to  $j=r=0$, vanishes. The   terms  with indices $j=0$, $r=1$, and $j=1$, $r=0$, provide the  'heavy-tailed' asymptotic behavior of the one-sided L\'{e}vy stable distributions. Consequently,  we can get see immediately that  the asymptotic behavior of $h_{\alpha_{1}, \alpha_{2}}(t, x)$ is proportional to $x^{-1-\min(\alpha_{1}, \alpha_{2})}$.

The first series in Eq. \eqref{8/07/14-1} can be expressed  as a  finite sum of  the generalized hypergeometric functions given in Eq. \eqref{26/08/14-6}. Indeed, let us consider, without loss of generality, the case of rational $\alpha_{1} > \alpha_{2}$, with $\alpha_{1} = \ulamek{l}{k}$ and $\alpha_{2} = \ulamek{p}{q}$, where $l$, $k$, $p$, and $q$ are integers. In this case,  Eq. \eqref{8/07/14-1} takes the form, 
\begin{align}
\label{14/07/14-1}
h_{\frac{l}{k}, \frac{p}{q}}(t, x) & = \frac{1}{\pi} \sum_{r, j = 0}^{\infty} \frac{(-t)^{r+j+1}}{r! j!} \frac{\Gamma(1 + \ulamek{l}{k} j + \ulamek{p}{q} r)}{x^{1 + \ulamek{l}{k} j + \ulamek{p}{q} r}} \sin\big[\pi\big(\ulamek{l}{k} j + \ulamek{p}{q} r\big)\big] \nonumber \\
& = \frac{1}{\pi} \sum_{r=0}^{\infty} \sum_{n=0}^{kp-1}\sum_{j=0}^{\infty} \frac{(-t)^{r+1+(n+kpj)}}{r! (n+kpj)!} \frac{\Gamma[1 + \ulamek{l}{k} (n + kpj) + \ulamek{p}{q} r]}{x^{1 + \ulamek{l}{k} (n + kpj) + \ulamek{p}{q} r}} \sin\big[\pi\big(\ulamek{l}{k} n + \ulamek{p}{q} r\big) + \pi l p j\big] \nonumber \\
& = \sum_{r=0}^{\infty} \sum_{n=0}^{kp-1} \frac{(-t)^{n+r}}{n!\, r!} \frac{x^{-1 -\ulamek{l}{k} n - \ulamek{p}{q} r}}{\Gamma\big(\!-\ulamek{l}{k} n - \ulamek{p}{q} r\big)} \sum_{j=0}^{\infty} \frac{(-1)^{kpj-lnj}}{j!} \left(\!\frac{t}{kp}\!\right)^{jkp} \left(\!\frac{lp}{x}\!\right)^{lp j} \nonumber \\
& \qquad \qquad \qquad \times \frac{(1)_{j} \prod_{s=0}^{lp-1} \big(\ulamek{1+s}{lp} + \ulamek{n}{kp} + \ulamek{r}{lq}\big)_{j}}{\prod_{s=0}^{kp-1} \big(\ulamek{1+j+s}{kp}\big)_{j}}.
\end{align}
To obtain  Eq. \eqref{14/07/14-1} we applied the Gauss-Legendre multiplication formula,  and the Euler's reflection formula. Now, using Eq. \eqref{26/08/14-6} we can represent the sum over $j$ in terms of the  generalized hypergeometric functions. Namely,  for $\ulamek{l}{k} \neq \ulamek{p}{q}$,  the function $h_{l/k, p/q}(t, x)$ can be written as  follows: 
\begin{align}
\label{14/07/14-2}
h_{\frac{l}{k}, \frac{p}{q}}(t, x) &= \sum_{r=0}^{\infty} \sum_{n=0}^{m_{1}-1} \frac{(-t)^{n+r}}{n!\, r!}\, \frac{\Gamma(1 + Mn + mr)}{x^{1+ M n + mr}} \sin(\pi M n + \pi mr) \nonumber \\
&\qquad \qquad  \times {_{1+lp}F_{m_{1}}}\left({1, \Delta(lp, 1 + M n + mr) \atop \Delta(m_{1}, 1+n)}; (-1)^{m_{1} - lp} \kappa\right),
\end{align}
with $\kappa = (lp)^{lp}/m_{1}^{m_{1}} (t^{1/M}/x)^{lp}$,  with $m,M,m_1,$ and $M_1$,  defined  in Eq.\eqref{26/08/14-10}. 
 Moreover, it turns out that, after applying Eq. (8.2.2.3) of \cite{APPrudnikov-v3}   to the finite sum in Eq. \eqref{14/07/14-2}, we can write 
\begin{equation}
\label{15/07/14-1}
h_{\frac{l}{k}, \frac{p}{q}}(t, x) = \frac{x^{-1} m_{1}\sqrt{M}}{(2\pi)^{\ulamek{m_{1}-lp}{2}}} \sum_{r=0}^{\infty} \frac{(-t)^{r}}{r!} \left(\frac{m_{1}}{t}\right)^{\ulamek{m}{M} r} G^{m_{1}, 0}_{lp, m_{1}}\left(\kappa\Big\vert {\Delta(lp, 0) \atop \Delta(m_{1}, \ulamek{m}{M} r)}\right),
\end{equation}
where $G^{m, n}_{p, q}$ is the Meijer $G$ function defined in Eq. \eqref{26/08/14-4}. 

The series representing  $h_{l/k, p/q}(t, x)$  given in Eq. \eqref{14/07/14-2} converges absolutely for  $x \geq 0$. The proof of this fact can be split into  two cases:  $x > 0$, and for $x = 0$. For $x > 0$, and fixed values of $l$, $k$, $p$, $q$, and $n=0, \ldots, m_{1}-1$, the convergence of the series $|h_{l/k, p/q}(t, x)|$ follows from the convergence of the ${_{1+lp}F_{m_{1}}}$  functions  which,  for a given $n$,  converges to a constant $C_{n}$ \cite{APPrudnikov-v3, NIST}.  Thus, we get
\begin{align}
\label{29/08/14-1}
|h_{\frac{l}{k}, \frac{p}{q}}(t, x)| &\leq \sum_{r=0}^{\infty}\sum_{n=0}^{m_{1}-1} \frac{t^{n+r}}{n! r!} \frac{\Gamma(1 + Mn + mr)}{x^{1+Mn + mr}} |\sin(\pi M n + \pi m r) C_{n}| \nonumber \\
& \leq \sum_{n=0}^{m_{1}-1} \frac{|C_{n}| t^{n}}{n!\, x^{M n}} \sum_{r=0}^{\infty} \frac{t^{r}}{r!} \frac{\Gamma(1 + M n + m r)}{x^{1 + mr}}.
\end{align}
An application of  the Cauchy ratio  test of convergence \cite{GBArfken05} to Eq. \eqref{29/08/14-1}  completes  the proof of convergence for  $|h_{l/k, p/q}(t, x)|$,  for $x > 0$. 

Let us now show that the series representing $h_{l/k, p/q}(t, x)$ converges absolutely at $x=0$. For this purpose  we will find the  asymptotic behavior of the ${_{1+lp}F_{m_{1}}}$  function in Eq. \eqref{14/07/14-2} for $\kappa \gg 1$, where the relation between $\kappa$ and $x$ is shown below Eq. \eqref{14/07/14-2}. It follows from  \cite{EWBarnes07}, that 
\begin{equation}
\label{14/07/14-3}
{_{1+ lp}F_{m_{1}}}(\ldots) \approx (-1)^{\frac{1}{2} + (M-1)n + mr} \frac{(lp)^{Mn + mr }}{m_{1}^{n}} \kappa^{\frac{1/2 - n + Mn + mr}{m_{1} - lp}} e^{- (m_{1}-lp) \kappa^{\ulamek{1}{m_{1} - lp}}},
\end{equation}
where, at the point $x = 0$, there exists   an essential singularity. That gives
\begin{equation}
\label{14/07/14-4}
h_{\frac{l}{k}, \frac{p}{q}}(t, x) \approx t^{\ulamek{1}{M}} \kappa^{1 + \ulamek{1}{2(m_{1} - lp)}} e^{-\big[m_{1} - lp + \big(\ulamek{t}{m_{1}}\big)^{m/M}\big] \kappa^{\ulamek{1}{m_{1}-lp}}} \sum_{n=0}^{m_{1}-1} (-1)^{M n} \sin(\pi M n) \kappa^{\ulamek{n}{m_{1}} + \ulamek{M-1}{m_{1} - lp}n}
\end{equation}
and, consequently, 
\begin{align}
\label{14/07/14-5}
|h_{\frac{l}{k}, \frac{p}{q}}(t, x)| &\leq t^{\ulamek{1}{M}} \kappa^{1 + \ulamek{1}{2(m_{1}-lp)}} e^{-\big[m_{1} - lp + \big(\ulamek{t}{m_{1}}\big)^{m/M}\big] \kappa^{\ulamek{1}{m_{1}-lp}}} \sum_{n=0}^{m_{1}-1} \kappa^{\big(\ulamek{1}{m_{1}} + \ulamek{M-1}{m_{1} - lp}\big)n} \nonumber \\
& \leq   t^{\ulamek{1}{M}} \kappa^{1 + \ulamek{1}{2(m_{1}-lp)}} e^{-\big[m_{1} - lp + \big(\ulamek{t}{m_{1}}\big)^{m/M}\big] \kappa^{\ulamek{1}{m_{1}-lp}}} m_{1} \kappa^{\big(\ulamek{1}{m_{1}} + \ulamek{M-1}{m_{1} - lp}\big)(m_{1}-1)} \nonumber \\
& = m_{1} t^{\ulamek{1}{M}} \kappa^{1 - \ulamek{1}{m_{1}} + \ulamek{1/2-M}{m_{1}-lp}} e^{-\big[m_{1} - lp + \big(\ulamek{t}{m_{1}}\big)^{m/M}\big] \kappa^{\ulamek{1}{m_{1}-lp}}}.
\end{align}
Taking into account the fact that, for $t > 0$, and fixed $l$, $k$, $p$, and $q$, such that $0 < l/k, p/q < 1$, and $l/k \neq p/q$, we can estimate Eq. \eqref{14/07/14-5}  as follows: 
\begin{equation}\label{14/07/14-6}
|h_{\frac{l}{k}, \frac{p}{q}}(t, x)| < B \kappa^{\ulamek{5}{2}} e^{- A \kappa},
\end{equation}
where $A$ and $B$ are positive constants. Now , from Eq. \eqref{14/07/14-6} it is easy to see that $|h_{l/k, p/q}(t, x)|\to 0$, for $x\to 0$. 

\bigskip

We will conclude this section with several concrete examples of explicit expressions for  bi-scale totally asymmetric L\'evy  densities, where we can show that some of our expressions can be reduced to expressions in terms of more classical special functions. \\

\noindent
 {\it Example 5.1.  The case   $\alpha_{1} = \alpha_{2} = 1/2$. }
Substituting the corresponding $v_{\alpha}(t, x)$  given in Eq. \eqref{26/09/2014-1a} into Eq. \eqref{25.06.14-1}, we have
\begin{align}
\label{19/09/2014-2a}
h_{\frac{1}{2}, \frac{1}{2}}(t, x) &= \frac{t^2}{4\pi} \int_{0}^{x} e^{-\ulamek{t^2}{4}\frac{x}{y(x-y)}} \frac{dy}{[y(x-y)]^{3/2}} = \frac{t^2}{2\pi} \int_{0}^{x/2} e^{-\ulamek{t^2}{4}\frac{x}{y(x-y)}} \frac{dy}{[y(x-y)]^{3/2}} \\ \label{19/09/2014-2b}
& = \frac{t^2}{\pi x^2} \int_{1}^{\infty} e^{-\ulamek{t^2}{x} z} \frac{dz}{\sqrt{z-1}} = \frac{t}{\sqrt{\pi} x^{3/2}} e^{-\ulamek{t^2}{x}} = g_{\frac{1}{2}}(x, 2t).
\end{align}
In Eq. \eqref{19/09/2014-2a} we changed the variable of integration  as follows: $z = 1/ [1-(1 - 2y/x)^2]$. In Eq. \eqref{19/09/2014-2b}  we also  employed  formula (3.362.1) on p. 344 of \cite{Gradshteyn}. The same result is obtained from Eq. \eqref{14/07/14-2} where, for $\alpha_{1} = \alpha_{2} = 1/2$, we employed the formulas  (7.11.3.3), (7.11.3.4) of \cite{APPrudnikov-v3} and (5.6.1.1) of \cite{APPrudnikov-v2}. \\

\noindent  
\textit{Example 5.2. The case $\alpha_{1} = 1/2$ and $\alpha_{2} = 1/3$.}
Using  Eqs. \eqref{26/09/2014-1a} and \eqref{26/09/2014-1b}, the Laplace convolution of $v_{\frac{1}{3}}(t, x)$, and $v_{\frac{2}{3}}(t, x)$, takes the form, 
\begin{align}
\label{26/09/2014-2}
h_{\frac{1}{2}, \frac{1}{3}}(t, x) & = \frac{t^{5/2}}{6 \pi^{3/2}} \int_{0}^{x} K_{\frac{1}{3}}\big(\ulamek{2}{\sqrt{x-y}} \ulamek{t^{3/2}}{3^{3/2}}\big) e^{-\frac{t^2}{4y}} \frac{dy}{[y(x-y)]^{3/2}} \nonumber \\
& = \frac{t^{5/2}}{6 \pi^{3/2} x^2} e^{-\frac{t^2}{4x}} \int_{0}^{\infty} u^{-3/2} (1+u) e^{-\frac{t^2}{4x} u} K_{\frac{1}{3}}\big(\sqrt{\ulamek{1+u}{x u}} \ulamek{2\, t^{3/2}}{3^{3/2}}\big) du
\end{align}
where $u = (x-y)/y$. Employing  the integer representation of $K_{\nu}(z)$ given in Eq. (8.432) on p. 917 in \cite{Gradshteyn}  
 we get

\begin{align}
\label{26/09/2014-3a}
h_{\frac{1}{2}, \frac{1}{3}}(t, x) & = \frac{t^{3} x^{-\frac{13}{6}}}{12\sqrt{3}\pi^{\frac{3}{2}}} e^{-\frac{t^2}{4x}} \int_{0}^{\infty} u^{-\frac{5}{3}} (1+u)^{\frac{7}{6}} e^{-\frac{t^2}{4x} u}\left[ \int_{0}^{\infty} \omega^{-\frac{4}{3}} e^{-\omega-\frac{t^3}{27 x \omega} \frac{1+u}{u}} d\omega\right] du \\ \label{26/09/2014-3b}
& = \frac{t^{3} x^{-\frac{13}{6}}}{12\sqrt{3}\pi^{\frac{3}{2}}} e^{-\frac{t^2}{4x}} \int_{0}^{\infty} u^{-\frac{5}{3}} (1+u)^{\frac{7}{6}} e^{-\frac{t^2}{4x} u}\left\{\int_{0}^{\infty} \omega^{-\frac{4}{3}} e^{-\omega} \sum_{r=0}^{\infty} \frac{[-\frac{t^3}{27 x \omega} \frac{1+u}{u}]^r}{r!} d\omega\right\} du \\ \label{26/09/2014-3c}
& = \frac{t^{3} x^{-\frac{13}{6}}}{12\sqrt{3}\pi^{\frac{3}{2}}} e^{-\frac{t^2}{4x}}  \sum_{r=0}^{\infty} \frac{(-\frac{t^3}{27 x})^r}{r!}  \int_{0}^{\infty} u^{-\frac{5}{3}-r} (1+u)^{\frac{7}{6} + r} e^{-\frac{t^2}{4 x} u} du \int_{0}^{\infty} \omega^{-\frac{4}{3}-r} e^{-\omega} d\omega\\  \label{26/09/2014-3d}
& = \frac{t^{3} x^{-\frac{13}{6}}}{12\sqrt{3}\pi^{\frac{3}{2}}} e^{-\frac{t^2}{4x}}  \sum_{r=0}^{\infty} \frac{(-\frac{t^3}{27 x})^r}{r!} \Gamma(-\ulamek{1}{3}-r) \Gamma(-\ulamek{2}{3}-r) \psi(-\ulamek{2}{3} - r; \ulamek{3}{2}; \ulamek{t^2}{4 x}) \\  \label{26/09/2014-3e}
& = \frac{2^{-\frac{1}{6}}\sqrt{\pi} t^2}{9\sqrt{3} x^{\frac{5}{3}}} e^{-\frac{t^2}{8 x}} \sum_{r=0}^{\infty} \frac{(-\frac{t^3}{3^3 2x})^r\, D_{\frac{7}{3} + 2 r}(\ulamek{t}{\sqrt{2 x}})}{r! \Gamma(\frac{4}{3}+r) \Gamma(\frac{5}{3}+r)} \\  \label{26/09/2014-3f}
&= \frac{\exp(-\frac{t^2}{8x})}{\sqrt{2\pi} x} \sum_{r=0}^{\infty} \frac{(-t)^{2+3r}}{(2+3r)!} (2 x)^{-\frac{2+3r}{3}}  D_{1 + \frac{2}{3}(2+3r)}(\ulamek{t}{\sqrt{2 x}}) \\  \label{26/09/2014-3g}
& = \frac{\exp(-\frac{t^2}{8x})}{\sqrt{2\pi} x} \sum_{n=0}^{\infty}  \frac{(-t)^{n}}{n!} (2 x)^{-\frac{n}{3}}  D_{1 + \frac{2}{3}n}(\ulamek{t}{\sqrt{2 x}})
\end{align}
To obtain Eq. \eqref{26/09/2014-3a} we used the series expansion of the exponential function, and thereafter we changed the order of integrals and summation. We also applied the definition of Tricomi's function (the confluent hypergeometric function of the second kind) $\psi(a; b; z)$ given in Eq. (9.211.4) on p. 1023 of \cite{Gradshteyn}. In Eq. \eqref{26/09/2014-3d} we employed Eq. \eqref{26/08/14-7} and formula (7.11.4.8) on p. 491 of \cite{APPrudnikov-v3}, whereas in Eq. \eqref{26/09/2014-3e} we used Eq. \eqref{26/08/14-8}. In Eq. \eqref{26/09/2014-3f} we changed the summation index from $2+ 3r$ to $n$. 
 
Note that Eq. \eqref{26/09/2014-3a} can be also obtained from Eq. \eqref{14/07/14-2}. Indeed,  using the formulas  (7.11.3.3) and (7.11.3.4) of \cite{APPrudnikov-v3} we get
\begin{align}
\label{29.06.14-3a}
h_{\frac{1}{2}, \frac{1}{3}}(t, x) & = \sum_{r=0}^{\infty} \frac{(-t)^{r}}{r!} \frac{x^{-\ulamek{r}{3}-1}}{\Gamma\big(\!-\!\ulamek{r}{3}\big)} {_{1}F_{1}}\left({1 + \frac{r}{3} \atop \frac{1}{2}} ; -\frac{t^{2}}{4x}\right)  + \sum_{r=0}^{\infty} \frac{(-t)^{r+1}}{r!} \frac{x^{-\ulamek{r}{3} - \ulamek{3}{2}}}{\Gamma\big(\!-\!\ulamek{r}{3}-\ulamek{1}{2}\big)} {_{1}F_{1}}\left({\frac{r}{3} + \frac{3}{2} \atop \frac{3}{2}} ; -\frac{t^{2}}{4x}\right) \nonumber \\
& = \frac{e^{-t^2/(8x)}}{\sqrt{2\pi} x} \sum_{r=0}^{\infty} \frac{(-t)^r}{r!} (2x)^{-r/3} D_{1+ \frac{2}{3} r}(\ulamek{t}{\sqrt{2x}}).
\end{align}

\newpage

\noindent
\textit{Example 5.3. The case $\alpha_{1} = 2/3$,  $\alpha_{2} = 1/3$.}
Here, the Laplace convolution defined in Eq. \eqref{25.06.14-1} is equal to
\begin{align}
\label{8/10/2014-1a}
h_{\frac{1}{3}, \frac{2}{3}}(t, x) & = \frac{t^{\frac{3}{2}}}{\sqrt{3} \pi^{\frac{3}{2}}} \int_{0}^{x} e^{-\frac{2t^3}{27 y^2}} W_{\frac{1}{2}, \frac{1}{6}}\big(\ulamek{4 t^3}{27 y^2}\big) K_{\frac{1}{3}}\big(\ulamek{2}{\sqrt{x-y}}\ulamek{t^{3/2}}{3^{3/2}}\big) \frac{d y}{y (x-y)^{3/2}}  \nonumber \\ 
& = \frac{[t/(3\pi)]^{3/2}}{2\sqrt{3}} \int_{1}^{\infty} \frac{\exp(-\frac{2t^3}{27 x^2} u)}{u^{\frac{1}{6}} (\sqrt{u}-1)^{\frac{5}{3}}}\, W_{\frac{1}{2}, \frac{1}{6}}\big(\ulamek{4 t^3}{27 x^2} u\big) K_{\frac{1}{3}}\big(\ulamek{2 t^{3/2}}{3^{3/2}} \ulamek{u^{1/4}}{(\sqrt{u} - 1)^{1/2}}\big) du \\ 
& = \frac{t^2 x^{-\frac{5}{3}}}{12\pi^{\frac{3}{2}}} \int_{1}^{\infty} \frac{\exp(-\frac{2t^3}{27 x^2} u)}{u^{\frac{1}{6}} (\sqrt{u}-1)^{\frac{5}{3}}}\, W_{\frac{1}{2}, \frac{1}{6}}\big(\ulamek{4 t^3}{27 x^2} u\big) \left[\int_{0}^{\infty} \omega^{-\frac{4}{3}} e^{-\omega - \frac{t^3}{27 x\omega} \frac{\sqrt{u}}{\sqrt{u}-1}} d\omega\right] du \nonumber \\ 
& = \frac{t^2 x^{-\frac{5}{3}}}{12\pi^{\frac{3}{2}}} \int_{1}^{\infty} \frac{\exp(-\frac{2t^3}{27 x^2} u)}{u^{\frac{1}{6}} (\sqrt{u}-1)^{\frac{5}{3}}}\, W_{\frac{1}{2}, \frac{1}{6}}\big(\ulamek{4 t^3}{27 x^2} u\big) \left\{\int_{0}^{\infty} e^{-\omega} \sum_{r=0}^{\infty} \frac{[-\ulamek{t^3}{27 x \omega}\ulamek{\sqrt{u}}{\sqrt{u}-1}]^r}{r!} \frac{d\omega}{\omega^{\frac{4}{3}} }\right\} du \nonumber \\ 
& = \frac{t^2 x^{-\frac{5}{3}}}{12\pi^{\frac{3}{2}}} \sum_{r=0}^{\infty} \frac{(-\frac{t^3}{27 x})^r}{r!} {\rm Int}(x, t) \int_{0}^{\infty} \omega^{-\frac{4}{3} - r} e^{-\omega}\, d\omega
 \nonumber \\ \label{8/10/2014-1b}
&= \frac{t^2 x^{-\frac{5}{3}}}{12\pi^{\frac{3}{2}}} \sum_{r=0}^{\infty} \frac{(-\frac{t^3}{27 x})^r}{r!} \Gamma(-\ulamek{1}{3} - r) {\rm Int}(x, t).  
%& = \frac{t^2 x^{-\frac{5}{3}}}{12\pi^{\frac{3}{2}}} \sum_{r=0}^{\infty} \frac{(-\frac{t^3}{27 x})^r}{r!} \Gamma(-\ulamek{1}{3}-r) \int_{1}^{\infty} u^{-\frac{1}{6}+\frac{r}{2}} (\sqrt{u}-1)^{-\frac{5}{3}+\frac{r}{2}} e^{-\frac{2 t^3}{27 x^2} u}\, W_{\frac{1}{2}, \frac{1}{6}}\big(\ulamek{4 t^3}{27 x^2} u\big) du 
\end{align}
Above, the  integral representation of $K_{\mu}(z)$, see  Eq. (8.432) on p. 917 in \cite{Gradshteyn},   was used. The auxiliary function ${\rm Int}(x, t)$ is defined  as follows: 
\begin{align}
\label{8/10/2014-2}
{\rm Int}(t, x) &= \int_{1}^{\infty} u^{-\frac{1}{6}+\frac{r}{2}} (\sqrt{u}-1)^{-\frac{5}{3}+\frac{r}{2}} e^{-\frac{2 t^3}{27 x^2} u}\, W_{\frac{1}{2}, \frac{1}{6}}\big(\ulamek{4 t^3}{27 x^2} u\big) du \nonumber \\
& = 2\sqrt{\frac{\pi}{3}}\, {_{2}F_{2}}\left({\ulamek{5}{6} + \ulamek{r}{2}, \ulamek{4}{3} + \ulamek{r}{2} \atop \ulamek{1}{3}, \ulamek{2}{3}}; -\frac{4 t^3}{27 x^2}\right) \nonumber \\
&- 2\sqrt{\frac{\pi}{3}} \frac{t}{x^{2/3}} \frac{\Gamma(-\ulamek{2}{3}-r)}{\Gamma(-\ulamek{4}{3}-r)}\, {_{2}F_{2}}\left({\ulamek{7}{6} + \ulamek{r}{2}, \ulamek{5}{3} + \ulamek{r}{2} \atop \ulamek{2}{3}, \ulamek{4}{3}}; -\frac{4 t^3}{27 x^2}\right) \nonumber \\
& + \sqrt{\frac{\pi}{3}} \frac{t^2}{x^{4/3}} \frac{\Gamma(-\ulamek{2}{3}-r)}{\Gamma(-2-r)}\, {_{2}F_{2}}\left({\ulamek{3}{2} + \ulamek{r}{2}, 2 + \ulamek{r}{2} \atop \ulamek{4}{3}, \ulamek{5}{3}}; -\frac{4 t^3}{27 x^2}\right),
\end{align}
where  Eq. (2.19.1.4) on p. 201 of \cite{APPrudnikov-v3} was employed. Substituting Eq. \eqref{8/10/2014-2} into Eq. \eqref{8/10/2014-1b}, using the Euler's reflection formula, the Gauss-Legendre multiplication formula, and changing the summation index as follows $n =  2 + 3r$, we get
\begin{align}
\label{15/07/14-4}
h_{\frac{1}{3}, \frac{2}{3}}(t, x) & = \sum_{n=0}^{\infty} \frac{(-t)^{n}}{n!} \frac{x^{-1-\ulamek{n}{3}}}{\Gamma(-\ulamek{n}{3})} {_{2}F_{2}}\left({\ulamek{1}{2} + \ulamek{n}{6}, 1 + \ulamek{n}{6} \atop \ulamek{1}{3}, \ulamek{2}{3}}; -\frac{4 t^{3}}{27 x^{2}}\right) \nonumber \\
& + \sum_{n=0}^{\infty} \frac{(-t)^{1+n}}{n!} \frac{x^{-\ulamek{5}{3}-\ulamek{n}{3}}}{\Gamma(-\ulamek{2}{3}-\ulamek{n}{3})} {_{2}F_{2}}\left({\ulamek{5}{6} + \ulamek{n}{6}, \ulamek{4}{3} + \ulamek{r}{6} \atop \ulamek{2}{3}, \ulamek{4}{3}}; -\frac{4 t^{3}}{27 x^{2}}\right) \nonumber \\
& + \sum_{n=0}^{\infty} \frac{(-t)^{2+n}}{2 n!} \frac{x^{-\ulamek{7}{3}-\ulamek{n}{3}}}{\Gamma(-\ulamek{4}{3}-\ulamek{n}{3})} {_{2}F_{2}}\left({\ulamek{7}{6} + \ulamek{n}{6}, \ulamek{5}{3} + \ulamek{n}{6} \atop \ulamek{4}{3}, \ulamek{5}{3}}; -\frac{4 t^{3}}{27 x^{2}}\right).
\end{align}

Finally, we would like to   point out that ${_{2}F_{2}}$'s cannot be expressed in terms  of  any other  special functions. And, of course, the explicit form of $h_{\frac{1}{3}, \frac{2}{3}}(x, t)$ can be obtained from Eq. \eqref{14/07/14-2} and Eq. \eqref{15/07/14-4}. In Fig. 5.2  we show  plots of the densities discussed in Examples 5.1-3. They were efficiently obtained via the symbolic manipulation platform {\it Mathematica}

\begin{figure}[!h]
\begin{center}
\includegraphics[scale=0.8]{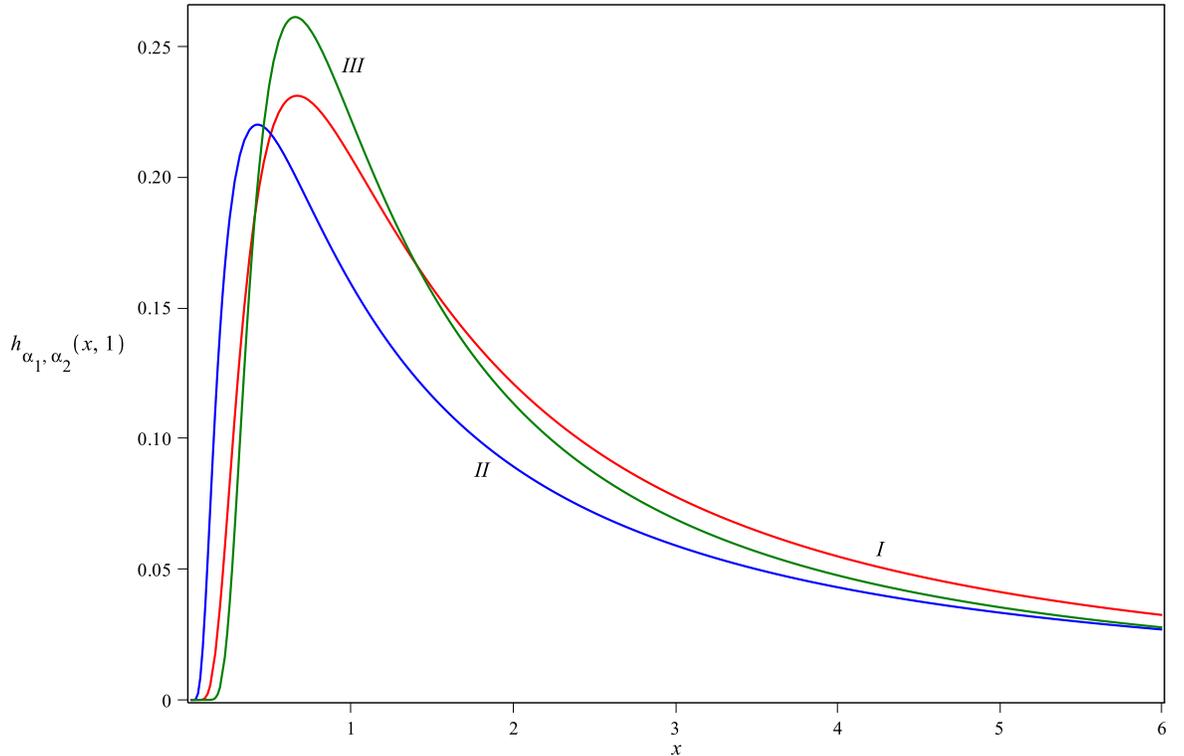}%{fig1.pdf}
\caption{\label{fig1} The plots  of $h_{\alpha_{1}, \alpha_{2}}(x, 1)$. Plot I (red):     $\alpha_{1} = \alpha_{2} = \ulamek{1}{2}$, see Eq. \eqref{19/09/2014-2a}); plot II (blue): $\alpha_{1} = \ulamek{1}{2}$,  $\alpha_{2} = \ulamek{1}{3}$, see Eq. \eqref{29.06.14-3a}; and  plot III (green): $\alpha_{1} = \ulamek{2}{3}$, $\alpha_{2} = \ulamek{1}{3}$, see Eq. \eqref{15/07/14-4}.}
\end{center}
\end{figure}

% ------------------------------------- Section 6 -------------------------------------------------

\section{Conclusions and comments; the  Stieltjes moment problem}

The  goal of this work was to show how the asymptotic behavior of solutions of certain nonlinear conservation laws can be explicitly  described in terms of some special functions which make the efficient computation of the related probability densities possible. In our future work we plan to expand this work to other types of asymptotic problems for other nonlinear evolution equations discussed in, e.g.,  \cite{KW2008},  and other papers referenced therein.

Here, we want to conclude with  an example of how the results obtained in  Section 5
 can be applied to the solution of the classical Stieltjes moment problem for special cases considered in Examples 5.1-3. 
Recall,  that the Stieltjes moment problem \cite{Akhiezer1965, KAPenson2009} can be formulated as follows: Find a  positive function $W(x)$ which satisfies the infinite set of equations, 
\begin{equation}
\label{10/10/2014-1}
\int_{0}^{\infty} x^{n} W(x) dx = \rho(n), \quad n=0, 1, 2, \ldots, 
\end{equation}
for a given moment sequence $\{\rho (n)\}$.

 From the practical point of view, it plays a special role in probability theory where the moment estimators can be usually conveniently  obtained, and the issue is whether   they determine the relevant probability distribution.  In general, the  solution to the Stieltjes moment problem can be unique or non-unique. The examples of non-unique solutions can be found, e.g., in \cite{KAPenson2009}. Below, we quote  one uniqueness criterion, and one non-uniqueness criterion to be used below.

 %\newpage 
 
 \begin{itemize}
\item[\textbf{C1}] {\it Carleman uniqueness criterion.} (see, .e.g.,   \cite{Akhiezer1965}).  
{If $\,S = \sum_{n=1}^{\infty} [\rho(n)]^{-\frac{1}{2n}}=\infty$, then $W(x)$ is uniquely determined by Eq. \eqref {10/10/2014-1}.}

\item[\textbf{C2}] {\it  Carleman non-uniqueness criterion, see, .e.g.,  {\rm\cite{AGPeaks01, AGut02}}.}     If $\sum_{n=1}^{\infty} [\rho(n)]^{-\frac{1}{2n}} < \infty$, and if there exists ${x' > 0}$ such that, for all $x > x'$, $W(x) > 0$, and $\psi(y) = -\ln[W(e^{y})]$ is convex in $y' > 0$, where $y' = \ln(x')$, then $W(x)$ is non-unique.
 
\end{itemize}

Let us begin with finding    the $\mu$-th moments of $h_{\alpha_{1}, \alpha_{2}}(x, t)$: 
\begin{equation}\label{31/07/14-4}
\rho_{\alpha_{1}, \alpha_{2}}(\mu) = \int_{0}^{\infty} x^{\mu} h_{\alpha_{1}, \alpha_{2}}(t, x) dx, \quad \mu\in\mathbb{R}.
\end{equation}
Using Eq. \eqref{15/07/14-1}, changing the variable $x$  to $y = x^{-lp}$, applying Eq. (2.24.2.1) in \cite{APPrudnikov-v3}, and the Gauss-Legendre multiplication formula (see Eq. \eqref{26/08/14-8}), we arrive at
\begin{align}
\label{15/07/14-2a}
\rho_{\alpha_{1}, \alpha_{2}}(\mu) %&= \frac{(2\pi)^{\ulamek{lp-m_{1}}{2}}}{\sqrt{M}} \sum_{r=0}^{\infty} \frac{(-t)^{r}}{r!} \left(\frac{m_{1}}{t}\!\right)^{\ulamek{m}{M}r} \int_{0}^{\infty} y^{-\ulamek{\mu}{lp}-1} G^{m_{1}, 0}_{lp, m_{1}}\left(\frac{t^{m_{1}}}{m_{1}^{m_{1}}} (lp)^{lp} y \Big\vert {\Delta(lp, 0) \atop \Delta(m_{1}, \ulamek{m}{M} r)}\right) \nonumber \\
& = \frac{1}{lp} \int_{0}^{\infty} y^{-\frac{\mu+1}{lp}-1} h_{\alpha_{1}, \alpha_{2}}(y^{-\frac{1}{lp}}, t) dy \\
& = \frac{1}{M} \sum_{r=0}^{\infty} \frac{(-1)^r}{r!}\, \frac{\Gamma(\frac{m r - \mu}{M})}{\Gamma(-\mu)}\, t^{(1-\frac{m}{M})r + \frac{\mu}{M}}, \quad t >0. \label{15/07/14-2b}
\end{align}
The absolute  convergence of the series in Eq. \eqref{15/07/14-2b} is easy to check through the Cauchy ratio test.  Eq. \eqref{15/07/14-2b} implies that the $\mu$th moment is finite for a given $r$,  and $\mu < m r$, whereas it is infinite for   $\mu\geq m r$.  The moment $\rho_{\alpha_{1}, \alpha_{2}}(\mu)$ is equal to one for $\mu =0$. The $\mu$-th moment of $h_{\alpha_{1}, \alpha_{2}}(x, t)$,   for $\mu = -lp n$, $n=0, 1, 2, \ldots$, gives rise to the Stieltjes moment problem  for which, from the comparison of Eq. \eqref{15/07/14-2a} with Eq. \eqref{10/10/2014-1},  $W(x)$ and $\rho(n)$  are of the form, 
\begin{equation}
\label{16/10/2014-1}
 W_{\alpha_{1}, \alpha_{2}}(t, x) = \frac{1}{lp} x^{-1-\frac{1}{lp}} h_{\alpha_{1}, \alpha_{2}}(t, x^{-\frac{1}{lp}}),  \quad \text{\rm and} \quad \rho(n) = \rho_{\alpha_{1}, \alpha_{2}}(-lp n).
\end{equation}
Now, we can employ  the above criteria \textbf{C1} and \textbf{C2}  to take a closer  look at the Stieltjes moment problem related to the Examples 5.1-3.

 In  Examples 5.1-2  functions $W_{\alpha_{1}, \alpha_{2}}(x)$ are unique. Indeed, for $\alpha_{1} = \alpha_{2} = 1/2$, we have $\rho(n) = 2 (2t)^{-2n} \Gamma(2n)/\Gamma(n)$,  which gives $S = 2 t \sum_{n=1}^{\infty} [2\Gamma(2n)/\Gamma(n)]^{-\frac{1}{2n}}$. Using the Stirling formula for the gamma function,  Eq. (8.327.1) on page 895 of \cite{Gradshteyn}, we get that the series $S \approx t \sqrt{e}/(2\sqrt{2}) \sum_{n=1}^{\infty} n^{-1/2}$; the latter series  is divergent. Thus, \textbf{C1} leads to the unique function $W_{\frac{1}{2}, \frac{1}{2}}(t, x) = t/(\sqrt{\pi x}) \exp(- x t^2)$. Similar considerations are valid  of the case of $\alpha_{1} = 1/2$ and $\alpha_{2} = 1/3$ which leads to the uniqueness of the appropriate function $W(t, x)$ in that case as well. 
 
 However, Example 5.3  leads to a non-unique Stieltjes moment problem. Indeed, for $\alpha_{1} = 2/3$, and $\alpha_{2} = 1/3$, we have $\rho(n) = 3 t^{-6n} \Gamma(6n)/\Gamma(2n)$, and the  series $S = t^3 \sum_{n=1}^{\infty} 3^{-\frac{1}{2n}} [\Gamma(6n)/\Gamma(2n)]^{-\frac{1}{2n}} \approx t^3 e^2/108 \sum_{n=1}^{\infty} 3^{-\frac{1}{4n}} n^{-2}$ is  convergent (we used  the Stirling formula  Eq. (8.327.1) on page 895 of \cite{Gradshteyn} here). The second condition in \textbf{C2} requires verification  of the sign of the second derivative of $\psi_{\frac{1}{3}, \frac{2}{3}}(t, x) = -\ln[W_{\frac{1}{3}, \frac{2}{3}}(t, e^{x})]$,  
 \begin{align}\label{15/01/2015-1}
\frac{d^2  \psi_{\frac{1}{3}, \frac{2}{3}}(t, x)}{d x^2} =& -\frac{d^2  \ln[W_{\frac{1}{3}, \frac{2}{3}}(t, e^{x})]}{d x^2}
= - u h_{\frac{1}{3}, \frac{2}{3}}(t, u) \frac{d}{d u} h_{\frac{1}{3}, \frac{2}{3}}(t, u)h_{\frac{1}{3}, \frac{2}{3}}(t, u) \nonumber \\
&  - u^2 h_{\frac{1}{3}, \frac{2}{3}}(t, u) \frac{d^2}{d u^2} + u^2 \left[\frac{d}{d u} h_{\frac{1}{3}, \frac{2}{3}}(t, u)\right]^2,
 \end{align}
for $u = \exp(-x/2)$. If Eq. \eqref{15/01/2015-1} is positive then  $\psi(y)$ is convex and $W_{\frac{1}{3}, \frac{2}{3}}(t, x)$ is non-unique. 
The positivity of Eq. \eqref{15/01/2015-1} with $h_{\frac{1}{3}, \frac{2}{3}}(t, u)$ given in Eq. \eqref{15/07/14-4} can be established analytically but for the purpose of this commentary section we are just showing the plot of Eq. \eqref{15/01/2015-1} in Fig. \ref{fig3}.
 \begin{figure}[!h]
\begin{center}
\includegraphics[scale=0.6]{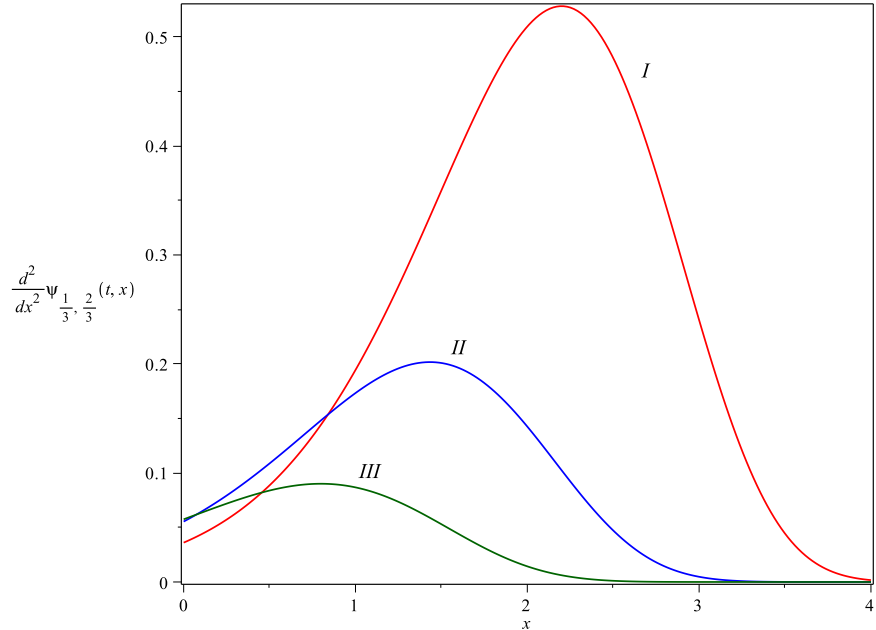}
\caption{\label{fig3} The plots of $\frac{d^2}{d y^2} \psi_{\frac{1}{3}, \frac{2}{3}}(t, x)$ for different values of $t$. Plot I (red): $t=0.8$, plot II (blue): $t=1$, and plot III (green): $t=1.2$. }
\end{center}
\end{figure}

\section*{Acknowledgment}
 
 K.~G\'{o}rska acknowledges support from the PHC Polonium, Campus France, project no. 28837QA and MNiSW under "Iuventus Plus 2015-2016" program no IP2014 013073. 
  
%------------------------------------------

\end{document}